\documentclass[a4paper,11pt]{article}
\pdfoutput=1 
\usepackage{jheppub} 
\usepackage{hyperref}
\usepackage{booktabs}
\usepackage[T1]{fontenc} 
\usepackage{siunitx} 
\usepackage{xcolor,graphicx}
\usepackage{dcolumn}
\usepackage{bm}
\usepackage[normalem]{ulem}
\usepackage{braket} 
\usepackage{amsmath}
\usepackage{cancel}
\usepackage{slashed}
\usepackage{multicol}
\usepackage[capitalise]{cleveref}
\usepackage{amssymb}
\DeclareFontFamily{U}{futm}{}
\DeclareFontShape{U}{futm}{m}{n}{
  <-> s * [.95] fourier-bb
  }{}
\DeclareSymbolFont{Ufutm}{U}{futm}{m}{n}
\DeclareMathAlphabet{\mathbbb}{U}{bbold}{m}{n}

\crefformat{subsubsection}{\S#2#1#3}

\counterwithin*{equation}{section}
\allowdisplaybreaks

\newcommand{\nnl}{\nonumber \\}

\newcommand{\cO}{\mathcal{O}}

\newcommand{\M}{{\mathcal M}}

\newcommand{\mpl}{M_{\mathrm Pl}}

\newcommand{\re}{{\mathrm{Re}} \,}

\usepackage{etoolbox,siunitx}
\robustify\bfseries

\begin{document}

\title{On-shell approach to scalar hair in spinning binaries}

\author{Adam Falkowski and}
\author{Panagiotis Marinellis}

\affiliation{Universit\'{e} Paris-Saclay, CNRS/IN2P3, IJCLab, 91405 Orsay, France}

\emailAdd{adam.falkowski@ijclab.in2p3.fr,
panagiotis.marinellis@ijclab.in2p3.fr 
          }


\abstract{
We propose an on-shell description of spinning binary systems in gravitational theories where compact objects display scalar hair.
The framework involves matter particles of arbitrary spin which, in addition to the minimal coupling to gravitons, are conformally coupled to a massless scalar mediating non-standard interactions. 
We use the unitary factorization techniques to derive the on-shell amplitudes relevant for emission of scalars and gravitons in matter scattering, paying attention to parametrize the freedom due to contact terms. 
Using the KMOC formalism, these amplitudes allow one to derive succinct expressions for the radiation waveforms at the leading post-Minkowskian order, together with the associated memory effects.
Furthermore, we compute the power emitted via gravitational and scalar radiation in hyperbolic encounters of compact objects. 
After a continuation to bound orbits, these are compared with  results obtained in specific scalar-tensor theories where black holes exhibit scalar hair, such as the scalar-Gauss-Bonnet or dynamical Chern-Simons theories. 
Finally, we identify possible deformations from the conformal coupling that can contribute to radiation observables at the same post-Newtonian order.
}
\maketitle

\section{Introduction} 

The dynamics of black hole (BH) binaries, particularly in the context of gravitational wave emission, has become a central focus in theoretical and observational astrophysics, spurred by the groundbreaking detections of LIGO and Virgo~\cite{LIGOScientific:2016aoc,LIGOScientific:2017vwq,KAGRA:2021vkt}. Accurately modeling these systems, especially in regimes beyond traditional post-Newtonian (PN) approximations (see e.g. Refs.~\cite{Levi:2018nxp,Porto:2016pyg,Blanchet:2013haa} for reviews of PN techniques), requires advanced methods for computing key observables such as waveforms, momentum transfer, and radiated energy. Among the most promising approaches are worldline methods (see e.g.~Refs.~\cite{WEFT1,WEFT2,WEFT3,WEFT4,WEFT5,WEFT6,Liu:2021zxr,Riva:2021vnj,Riva:2022fru} and~\cite{WQFT1,WQFT2,WQFT3,Ben-Shahar:2023djm,Jakobsen:2021smu,Jakobsen:2021lvp} for two alternative formalisms) and modern scattering amplitude techniques~\cite{Solon1,Bern1,Bern2,Bern:2024adl,Bern:2020buy,Exp1,Exp2,Heavymass1,Kosower:2018adc,Maybee:2019jus,Cristofoli:2021vyo,Eikonal1,Eikonal2,Eikonal3,Eikonal4,EikonalKMOC,Luna:2023uwd,Aoude:2022trd,Brandhuber:2023hhy,Elkhidir:2023dco,Herderschee:2023fxh,Georgoudis:2023lgf,Caron-Huot:2023vxl,DeAngelis:2023lvf, Aoude:2023dui, Brandhuber:2023hhl, Bohnenblust:2023qmy}, both of which offer complementary strengths in addressing the complexities of BH interactions. 
Worldline methods provide an elegant framework by modeling BHs as effective point particles tracing worldlines in curved spacetime. These techniques naturally capture both the conservative and dissipative aspects of gravitational dynamics and offer a systematic way to compute observables through a perturbative expansion in the gravitational coupling. 
In parallel, scattering amplitude methods, originating in high-energy physics, have recently been adapted to gravitational systems with great success.
The Kosower-Maybee-O'Connell (KMOC) formalism has been particularly instrumental in this endeavour~\cite{Kosower:2018adc,Maybee:2019jus,Cristofoli:2021vyo}.
By leveraging the symmetries inherent in scattering processes, these techniques simplify the computation of complex gravitational interactions. 
The on-shell amplitude methods together with the spinor helicity variables~\cite{Arkani-Hamed:2017jhn}, which reveal the simplicity of gauge theory and gravity interactions, have been particularly influential in enabling efficient calculations in the post-Minkowskian (PM) framework. 
These PM results can be important not only for systems with high velocities but also with large eccentricities at fixed periastron distances~\cite{Khalil:2022ylj}. 
In fact, they have already been implemented in the context of standard general relativity (GR) to derive precise gravitational waveform templates for bound orbits~\cite{Buonanno:2024byg}. 
All in all, the amplitude-based approach is now playing a central role in improving the precision of gravitational waveform predictions, offering a novel and powerful approach to the two-body problem in GR.

The next generation of gravitational wave interferometers, such as LISA~\cite{LISA:2017pwj}, the Einstein Telescope~\cite{Sathyaprakash:2012jk,Maggiore:2019uih} and Cosmic Explorer~ \cite{Evans:2021gyd}, will offer higher precision, allowing for more accurate tests of general relativity. 
This motivates the consideration of beyond-GR effects in this paper, as we aim to explore potential deviations that could be detected with these advanced instruments. 
Moreover, there is also theoretical interest into how these effects manifest themselves in a Quantum Field Theory (QFT) framework. 
Recent works have been exploring these avenues by using the worldline~\cite{Bhattacharyya:2024aeq, Bhattacharyya:2024kxj} and amplitude methods~\cite{Davis:2023zqv,Brandhuber:2024bnz,Falkowski:2024bgb,Brandhuber:2024qdn}.
In this paper we focus on the so-called scalar-tensor theories, where interactions are mediated by a massless scaler $\phi$, in addition to the spin-2 graviton of GR. 
In such a framework, we study waveforms of gravitational and scalar radiation emitted in hyperbolic collisions of arbitrary spinning bodies with scalar hair. 
In contrast with Refs.~\cite{Bhattacharyya:2024aeq, Bhattacharyya:2024kxj,Davis:2023zqv} we will apply modern amplitude techniques and the KMOC formalism to describe the effect of the scalar hair, which was not attempted in the previous literature.
In practice, the scalar hair implies a certain coupling of the scalar to the matter particles representing macroscopic bodies. 
The present work, apart from being a natural extension of Ref.~\cite{Falkowski:2024bgb} where the scalar was assumed not to couple directly to matter, provides a general framework that can be applied to any scalar-tensor theories where compact objects acquire scalar hair as described below. 
This framework will also be applicable to scalar-tensor theories of the Brans-Dicke class \cite{Jordan1955-JORSUW,Fierz:1956zz,PhysRev.124.925,Lang:2014mra,Lang:2013fna,Lang:2014osa}. We will support this claim by comparing our results to classical calculations of emitted power in scalar-tensor theories, whenever such results are available. 

Hairy objects are a frequent occurrence in modified theories of gravity admitting non-GR solutions~\cite{Charmousis:2014zaa,Herdeiro:2015gia,Delgado:2016jxq,Pacilio:2018gom,R:2022tqa,Bakopoulos:2023fmv}. There are various examples in theories containing at least one massless degree of freedom as for example the case of scalar Gauss-Bonnet (SGB)~\cite{Kanti:1995vq, Maeda:2009uy,Sotiriou:2013qea, Sotiriou:2014pfa} or dynamical Chern-Simons (DCS) theories~\cite{Alexander:2009tp, Yunes:2009hc}. 
In the class of theories which also admit the known GR solutions, there are also various processes which can lead to compact objects becoming \emph{scalarized}, i.e. acquiring scalar hair through some mechanism (see Ref.~\cite{Doneva:2022ewd} for a review on models of scalarization). For a review on BHs with scalar hair see for example~\cite{Herdeiro:2015waa}. 
In more detail, a compact object can acquire different kinds of scalar charges depending on the
field configuration in the far zone. 
In that regime the scalar field can be expanded in powers of $1/r$, where $r$ is an appropriate measure of distance from the compact object. 
In this expansion, we call the coefficient of the leading $1/r$ term "monopole scalar charge" (or "monopole hair"). The rest of the coefficients of the $1/r^2$, $1/r^3$, etc.~\footnote{
Note that, within this characterization,  only the leading non-vanishing scalar charge has an invariant meaning.} terms are naturally called "dipole hair", "quadruple hair", etc.
Note that scalar hair can be either secondary, which is the case of the SGB/DCS theories, being a function of the properties of the BH and the coupling of the theory considered\footnote{In the examples of SGB/DCS, the scalar charge $q=q(m,S,Q,\frac{\hat{\alpha}}{\Lambda^2})$, where $m,S,Q$ are respectively the mass, spin and electric charge of the BH~\cite{Julie:2019sab,Yagi:2012vf}.},  or primary, i.e. a global charge distinct from mass, spin or electric charge. 
In this article we will primarily focus on the case of compact objects with "monopole hair" since the effects' of the rest of the terms are highly suppressed compared to that one. 
We will also discuss the possibility to describe the effects of a "dipole scalar charge" within the amplitude-based formalism. 
Lastly, we identify generic deviations from the monopole-like coupling of the scalars to matter,  which contribute to classical observables at the same PN order. 
We remain, though, agnostic about the source of these modifications.

Experimental signatures from such objects have been a subject of research for some time now with various works focused on theories where BHs and/or neutron stars (NSs) can have scalar hair~\cite{Yagi:2011xp,Yagi:2013mbt,Yagi:2012vf,Okounkova:2017yby,Julie:2018lfp,Khalil:2018aaj,Witek:2018dmd,Loutrel:2018ydv,Julie:2019sab,Shiralilou:2021mfl,Perkins:2021mhb,Yang:2022ifo,Corman:2022xqg,Julie:2024fwy,Antoniou:2024hlf,Bernard:2022noq,Lang:2014osa}. 
One main phenomenological impact of the monopole scalar hair enters via dipole scalar radiation, leading to a potentially observable anomalous energy loss of the binary system.  
Most of the studies so far make use of the PN scheme or numerical simulations to study observables by evolving the equations of motion for a binary system, or by calculating the modifications to the quasi-normal modes of the BHs. 
Results from these approaches have already been implemented to provide rigid bounds on the couplings of scalar-tensor theories: in particular recent work using the existing gravitational wave data has provided the most stringent bound up to date for SGB~\cite{Julie:2024fwy}, $\sqrt{\alpha_{\rm SGB}} \lesssim 0.31$ km following the conventions of that paper (see also~\cite{Silva:2022srr} for bounds on DCS from gravitational wave observations\footnote{%
More stringent bounds for DCS were derived from neutron star observations~\cite{Silva:2020acr} and lead to $\sqrt{\alpha_{\rm DCS}} \lesssim 8.5$ km in their conventions.}).
Nevertheless, this can be an extremely demanding task because there can be very complicated systems of equations to be solved simultaneously, and  because it is often very difficult to find (even approximate) solutions for BHs or NSs in these theories. 
On the other hand, amplitude methods can bypass the aforementioned problems and provide precise predictions for binaries in modified GR at all orders in velocity and any requested order in spin. 
We note, however, that the amplitudes are designed to model scattering events. 
In order to map the observables to the (phenomenologically more relevant) bound case one should use a dictionary, such as the one presented in Refs.~\cite{Kalin:2019rwq,Kalin:2019inp,Cho:2021arx}.
We also remark that certain aspects of that dictionary are still subject of ongoing research, notably in the context of the static contributions to the waveform~\cite{Adamo:2024oxy}. 

This paper is organized as follows. 
In~\cref{sec:scalarization} we discuss the on-shell amplitude description of compact objects with monopole scalar hair. 
We consider matter particles with a conformal coupling to the massless scalar, and derive the classical limit of the corresponding 3-point amplitude for any matter spin. 
From that starting point, using unitarity, we build 5-point amplitudes describing scalar or graviton emission  in matter scattering, paying attention to the contribution of contact terms in the intermediate 4-point amplitudes. 
Through the KMOC approach, the 5-point amplitudes, or more precisely their residues, serve in~\cref{sec:waveforms} as an input to calculate the waveforms for scalar and gravitational radiation in hyperbolic matter scattering. 
We also calculate the emitted power, tracing up to linear corrections in spin and up to quadratic corrections proportional to the scalar monopole charge.
We argue that these results can be connected to energy loss in quasi-circular bound-state systems.  
We show that the energy loss formula obtained in the context of classical  scalar Gauss-Bonnet theories can be reproduced in the amplitude formalism by taking the non-relativistic limit of the waveforms and using the Kepler law. In~\cref{sec:non-conformal} we discuss generic classical non-conformal couplings between the scalar and matter and study their leading effects. 
We conclude in \cref{sec:conclusions}. 
Our conventions are briefly summarized in \cref{app:CON}.  
In \cref{app:IntRes} we review the method to calculate the integrals occurring in the derivations of the waveforms. 
Some additional details about the waveforms and amplitudes are relegated to \cref{app:MWF,app:higheramps}. 
Finally, \cref{app:CONTACT} discussess an algorithm to generate the relevant contact terms at any order in spin.

\section{Spinning compact objects with scalar hair on-shell} 
\label{sec:scalarization}

In this section we present our framework to describe BHs endowed with scalar hair. 
The QFT techniques to calculate classical GR observables rely on representing compact objects by spinning point-like matter particles coupled to gravitons,  where the coupling is minimal in the case of BHs.    
Scalar hair implies that matter also couples to the massless scalar $\phi$.  
In particular, we will argue that the monopole hair corresponds to matter coupled to an effective metric $\tilde g_{\mu \nu}$ with a specific dependence on $\phi$. 
The most general effective metric compatible with causality is the Bekenstein metric~\cite{Bekenstein:1992pj}, which can be written in the following form:
\begin{align}
\label{eq:CC_tildeg}
    \tilde{g}_{\mu \nu} = \exp \Big [ C\big ( \frac{\phi}{\mpl} \big ) \Big ] g_{\mu \nu} + D\big ( \frac{\phi}{\mpl} \big ) \frac{D_\mu \phi D_\nu \phi}{\mpl^2 \Lambda^2},
\end{align}
where $\Lambda$ is some dimensionful parameter controlling the derivative expansion of the scalar interactions. For $D\big ( \frac{\phi}{\mpl} \big ) = 0 $ the transformation $g_{\mu\nu} \to \tilde g_{\mu\nu}$ is called \emph{conformal}, while for $C\big ( \frac{\phi}{\mpl} \big ) = 0$ one has a \emph{purely disformal} transformation.
In this article we turn off the disformal factor $D$ since its effects are further suppressed by the additional energy scale $\Lambda$. 
In the following we will focus on the couplings between matter and the scalar $\phi$ induced by the conformal factor $C$.
For our purpose it is enough to approximate 
\begin{align}
    C\big ( \frac{\phi}{\mpl} \big ) \approx 1 + c \frac{\phi}{\mpl} 
, \end{align}
with an unknown constant $c$ characterizing the interaction strength and ultimately translating into scalar monopole charge. 
In the on-shell approach, $\cO(\phi^2)$ and higher interactions terms in the Lagrangian are not needed for constructing amplitudes.  

\subsection{3-point amplitudes}

We first work out the on-shell 3-point amplitude between matter and the scalar $\phi$ interacting  via the effective metric in~\cref{eq:CC_tildeg}.
To this end one needs to perform the $g_{\mu\nu} \to \tilde g_{\mu\nu}$ transformation in the Lagrangian, and work out the relevant cubic interaction terms between the scalar and matter. 
A useful shortcut is provided by the observation~\cite{Sotiriou:2008rp,DelGrosso:2023trq}
(for a worldline perspective on that, see e.g. Refs.~\cite{1975ApJ...196L..59E, Damour:1992we,Damour:1995kt}\footnote{In the GR literature this approach is often referred to as a \emph{skeletonization} procedure.})
that the conformal transformation is equivalent to rescaling the matter mass term as  
\begin{align}
\label{eq:CC_mtoexpcm}
m \to e^{C/2} m \approx  \bigg ( 1 + c \frac{\phi}{2 \mpl} \bigg ) m 
    .  \end{align}
We have verified this for matter of up to  spin two.
Given \cref{eq:CC_mtoexpcm}, it is straightforward to calculate the on-shell 3-point amplitude for any matter spin in terms of spinor helicity variables. 
For bosonic and fermionic matter we find the compact expression 

\begin{align}
\label{Amp_3_bos}
\M \big [ 1_{\Phi_n}, 2_{\Phi_n}, 3_\phi \big ] = & 
 -  \frac{c_n}{\mpl} \frac{\langle \bm 2 \bm 1 \rangle^{S_n}  [\bm 2 \bm 1]^{S_n}}{m_n^{2S_n-2}}  
 , & \quad S_n =  k
, \nnl 
\M\big [ 1_{\Phi_n}, 2_{\bar {\Phi}_n}, 3_\phi \big ] =  & 
- \frac{c_n}{ \mpl} \frac{ \langle \bm 2 \bm 1 \rangle^{S_n-1/2} [ \bm 2 \bm 1]^{S_n-1/2} \Big ( \langle \bm 2 \bm 1 \rangle  + [ \bm 2 \bm 1] \Big ) } {m_n^{2S_n-2}}
, & \quad S_n = k+ 1/2 
, \end{align}
where $k$ is a positive integer, and the subscript $n$ labels different matter particles with $m_n$ and $c_n$ their associated masses and conformal couplings as in~\cref{eq:CC_mtoexpcm}. 
Here, $\Phi_n$ and $\bar \Phi_n$ represent incoming (outgoing) matter.
Each is associated with $2S_n$ upper (lower) little group indices, which are carried by the spinors 
$| \bm 1 \rangle$ ($| \bm 2 \rangle$) and which encode the polarization.
The little group indices are suppressed most of the time to reduce clutter. 
We notice the simple pattern of the amplitudes for particles of different spin when written in the spinor-helicity language. 
In the classical limit both the bosonic and fermionic amplitudes converge to a unique answer:
\begin{align}
    \M^{\rm cl} \big [1_{\Phi_n}, 2_{\bar \Phi_n}, 3_\phi \big ] = 
    - c_n \frac{m_n^2}{\mpl} 
    \label{Amp_phi-Phi-Bar{Phi}}
. \end{align}
The above is implicitly multiplied by $\mathbbb{1}^{2S} \equiv \delta^{I_1}_{J_1} \dots \delta^{I_{2S}}_{J_{2S}}$ with the upper and lower little group indices separately symmetrized. 
To arrive at the above one has to simultaneously take the $S_n \to \infty$, $\hbar \to 0$ limits and express the massive spinors in terms of one another:
    \begin{align}
&|\bm 2 \rangle = -  | \bm 1 \rangle - \frac{|p_3|\bm 1]}{2m}  
, \qquad 
| \bm 2 ] =  |\bm 1] + \frac{|p_3|\bm 1\rangle}{2m}   . 
    \end{align}
In the classical limit, the spin dependence becomes trivial for monopole charge couplings. 
Moreover, the interaction survives in the limit where the scalar momentum goes to zero, $p_3 \to 0$, 
unlike the scalar interactions considered in Ref.~\cite{Falkowski:2024bgb}. 
For BHs with scalar hair in the SGB framework one has 
$c_n \sim \frac{\alpha \mpl^4}{\Lambda^2 m_{n}^2} +\mathcal{O}(\alpha^2, a) $, $\alpha$ is the magnitude of the coupling of the scalar to the Gauss-Bonnet invariant~\cite{Julie:2019sab} and $a^\mu$ is the classical spin vector associated with the BH. 
For this reason, we will treat $c_n$ as small parameters, and only trace  their leading effects in observables.  
More precisely, we will be interested in $\cO(c_n^2)$ corrections to power emitted during scattering, which implies tracing up to $\cO(c_n)$ effects in the scalar waveforms, and up to $\cO(c_n^2)$ effects in the gravitational waveforms.  

To evaluate  4- and 5-point amplitudes relevant for the waveform calculations we will also need the gravitational couplings of matter and the scalar.
We  assume minimal coupling of matter with gravity, translating to the following 3-point on-shell amplitudes~\cite{Arkani-Hamed:2017jhn}:
\begin{align}
    \M \big [1_{\Phi_n} 2_{\bar \Phi_n} 3_h^- \big ] = & 
    -  {\langle 3 |p_1|\tilde{\zeta}]^2  \over \mpl [3 \tilde{\zeta}]^2}   \frac{[\bm 2 \bm 1]^{2S_n}}{m_n^{2S_n}}
, \nnl 
\M  \big [ 1_{\Phi_n} 2_{\bar \Phi_n}  3_h^+ \big ] = & 
-{\langle \zeta|p_1|3]^2  \over \mpl \langle 3 \zeta \rangle^2} 
  \frac{\langle \bm 2 \bm 1 \rangle^{2S_n}}{ m_n^{2S_n}} .
\label{Amp_3ptMinCoupling}
\end{align}
In the classical limit, 
\begin{align}
    \M^{\rm cl} \big [1_{\Phi_n} 2_{\bar \Phi_n} 3_h^- \big ] = & 
    -  {\langle 3 |p_1|\tilde{\zeta}]^2  \over
    \mpl [3 \tilde{\zeta}]^2}  
    \exp \big [ -  p_3 a_n \big ]
, \nnl 
\M^{\rm cl} \big [ 1_{\Phi_n} 2_{\bar \Phi_n}  3_h^+ \big ] = & 
-{\langle \zeta|p_1|3]^2  \over 
\mpl \langle 3 \zeta \rangle^2} 
\exp \big [ p_3 a_n \big ]
\label{Amp_3ptMinCoupling-classical}
, \end{align}
where $a_n$ is the spin vector associated with $\Phi_n$ defined as\footnote{
More generally, if the matter field $\Phi_n$ appears at the $k$-th position in the amplitude, 
one should replace  $|\boldsymbol{1}]$ 
and $|\boldsymbol{1} \rangle $ in~\cref{eq:SPD_spinvector} 
by  $|\boldsymbol{k}]$ and $|\boldsymbol{k} \rangle $.}
 \begin{align} 
   \label{eq:SPD_spinvector}
a_n^\mu = 
- S_n {\langle \bm 1 |\sigma^\mu| \bm 1] 
+ [\bm 1 |\bar \sigma^\mu| \bm 1 \rangle \over 2m_n^2} \mathbbb{1}^{2S_n-1}
.   \end{align} 
The amplitude in~\cref{Amp_3ptMinCoupling-classical} in the limit $a_n \to 0$ also describes the coupling of the massless scalar $\phi$ to gravity.
For the holomorphic and anti-holomorphic 3-point kinematics, respectively, it can be rewritten in the familiar form: 
\begin{align}
    \M \big [1^-_{h}, 2_\phi, 3_\phi ] =& -\frac{1}{ M_{Pl}} \bigg ( \frac{\langle 12\rangle \langle 1 3 \rangle }{\langle 2 3 \rangle } \bigg )^2  
    , \nnl
     \M \big [1^+_{h}, 2_\phi, 3_\phi ]=& -\frac{1}{ M_{Pl}} \bigg ( \frac{[ 12 ] [ 1 3 ] }{[ 2 3 ] } \bigg )^2 
     \label{Amp_h-phi-phi}
. \end{align}

If we specify our analysis to the scalar-tensor theories considered in Ref.~\cite{Falkowski:2024bgb} the scalar has also a helicity flipping coupling to two gravitons proportional to the SGB/DCS coupling $\hat{\alpha}$. However, as we will later see, the contributions to the waveforms computed in that paper are much more suppressed than the ones that come solely from the conformal coupling. For this reason, in the present paper we ignore the effects of that coupling on the waveforms.

\subsection{Four-point amplitudes} 
\label{sec:Four-point_amplitudes}

Next we move on to discuss the four-point amplitudes. 
We split the contributions into a \emph{unitary} and a \emph{contact} part:
\begin{align}
    \M_4 =  \M_{4,U} +  \M_{4,C},
\end{align}
where the $\M_{4,U}$ has all the physical poles required by unitary factorization, and $\M_{4,C}$ does not have  any physical poles. 
We will  later expand each one of the two amplitudes in powers of the spin:
\begin{align}
    \M_{4,X} = \sum_{n=0}^{2S} \M^{(n)}_{4,X} ,
\end{align}
with $X= \{U,C\}$.  
From now on we restrict our attention to the leading order effects in the scalar charges, given that we expect the conformal coupling to be small, $c \ll 1$. We will quote only the classical limit of the relevant amplitudes in the main text. For the full four-point quantum amplitudes we refer the readers to \cref{app:Full_Four-points}.

The first amplitude to consider is the scattering of a massless scalars off a massive particle. 
In this case there are actually two contributions to consider: one coming from  gluing  the amplitudes in~\cref{Amp_3ptMinCoupling} with then one in~\cref{Amp_h-phi-phi}, and another from gluing two of the three-point amplitudes in~\cref{Amp_phi-Phi-Bar{Phi}}. 
However, the latter is proportional to $c_{n}^2$ and is neglected. The result for the case we are interested in reads:
\begin{align}
 \M^{\rm cl}_{U}\big [1_{\Phi_n}, 2_{\bar{\Phi}_n}, 3_\phi, 4_\phi\big ]  &=  \frac{t-m_n^2}{\mpl^2 s} \bigg \{  
  -(t-m_n^2) \cosh (q \cdot a_n) 
  \nnl & 
  + 2i \varepsilon_{\mu \nu \rho \sigma} p_1^\mu p_3^\nu p_4^\rho a_n^\sigma \frac{\sinh (q \cdot a_n)}{q \cdot a_n} \bigg \} + \cO(c_n^2)
 \label{Amp_Phi-bar{Phi}-phi-phi} 
, \end{align}
where $q^\mu \equiv p_1^\mu + p_2^\mu$ and the Mandelstam invariants are defined as 
$s=(p_1 +p_2)^2 , t=(p_1 +p_3)^2 , u=(p_1 +p_4)^2$.
Notice that the amplitude scales as $\sim \hbar^0$ under the classical (or soft) scaling $p_3 \sim p_4 \sim q \sim \hbar^1$, $a_n \sim \hbar^{-1}$. 
Another comment is that the Bose symmetry under the $3 \leftrightarrow 4$ exchange is obscured  in~\cref{Amp_Phi-bar{Phi}-phi-phi} because  $t-m^2 \approx - (u-m^2)$ in the classical limit. 
To arrive at \cref{Amp_Phi-bar{Phi}-phi-phi} one has to expand the spinors as follows~\cite{Aoude:2020onz, Cangemi:2023ysz}:
    \begin{align}
&|\bm 2 \rangle = - \frac{1}{\sqrt{1-\frac{q^2}{4m^2}}} \bigg [ | \bm 1 \rangle - \frac{|q|\bm 1]}{2m}   \bigg ] 
, \nnl
&| \bm 2 ] = \frac{1}{\sqrt{1-\frac{q^2}{4m^2}}}  \bigg [ |\bm 1] 
- \frac{|q|\bm 1\rangle}{2m}  \bigg ] 
.     \end{align}
Moving to the scattering of 2 massive particles off a scalar and a graviton, we have a contribution proportional to the monopole charge entering in all three factorization channels.
We find
\begin{align}
    \M^{ \rm cl}_{U}\big [1_{\Phi_n}, 2_{\bar{\Phi}_n}, 3_\phi, 4_h^- \big ] = &
    -\frac{c_n m_n^2 }{\mpl^2}  \frac{\langle 4 |p_3 p_1|4 \rangle^2 }{s (t-m_n^2)^2 } 
    , \nnl
    \M^{ \rm cl}_{U}\big [1_{\Phi_n}, 2_{\bar{\Phi}_n}, 3_\phi, 4_h^+ \big ] =  &  -
    \frac{ c_n m_n^2 }{\mpl^2}  \frac{[ 4 |p_3 p_1|4 ]^2}{s (t-m_n^2)^2 } 
    . \label{Amp_Phi-bar{Phi}-phi-h}
\end{align}
Once again, the amplitude is $\sim \hbar^0$ under classical scaling.

One could also consider the corrections to the gravi-Compton amplitude coming from the inclusion of the SGB/DCS and conformal coupling. 
However, this contribution goes as $\Delta\M_{\rm Compton} \sim \hat{\alpha} c_n$, which we neglect in observables.\footnote{Nevertheless, that amplitude can be found in~\cref{app:Subleading_Amp_Res_SGB/DCS}, along with for a more precise argument on why this contribution is subleading. 
We also discuss classical contact deformations to the aforementioned amplitude in~\cref{app:Contact_2mass_2grav}.}

We turn to discussing the contact terms. As mentioned earlier,  in this section we are interested in tracing up to linear-in-spin effects in the emitted power, therefore we focus on the contact terms that might contribute at that order. 
We have verified that the contact deformations discussed below are the only ones generated at each respective order by building an ansatz to generate the complete set of classical contact terms at all spin orders, see~\cref{app:CONTACT}. 
This follows similar lines as in Appendix~B of Ref.~\cite{Falkowski:2024bgb}, except now we are interested in contact terms scaling as $\hbar^0$ in the classical limit, and we also expand the discussion to include contact terms for the classical scattering of scalars and gravitons off a massive particle. In the following we will just quote the results from this analysis which, for low spin orders,  can also be easily obtained by direct inspection. 

We first focus on 
$\M_C^{\rm cl}\big [1_{\Phi}, 2_{\bar{\Phi}}, 3_\phi, 4_\phi\big ]$.
In this case the classical contact term actually appears already at the zeroth spin order. This is in contrast to the situation for classical electromagnetic or gravitational (gravi-)Compton amplitudes. 
The reason is that the amplitude at hand involves scalars rather than photons and gravitons, which do not have  helicity weights, and thus a constant contact terms is allowed. 

We parametrize this contact term as
\begin{align}
     \M^{\rm cl, (0)}_{C}\big [1_{\Phi_n}, 2_{\bar{\Phi}_n}, 3_\phi, 4_\phi\big ] = 
     \frac{C_n^{(0)} m_n^2}{\mpl^2} 
     \label{Amp_Phi-bar{Phi}-phi-phi_Contact0} 
, \end{align}
where $C_1^{(0)}$ is a real coefficient. 
At the linear order in spin we have  
\begin{align}
\M^{\rm cl, (1)}_{C}\big [1_{\Phi_n},2_{\bar{\Phi}_n}, 3_\phi,4_\phi\big ] = 
     - i \frac{C_n^{(1)} m_n^2}{\mpl^2} (q a_n)  
     \label{Amp_Phi-bar{Phi}-phi-phi_Contact1} 
. \end{align}
Naively, one could also try to include an absorptive contact term at this order, 
$ \M^{\rm cl, (1)}_{C} \sim (p_1 q) |a_n|$, but that is vanishing in the classical limit because $p_1 q \sim \hbar^2$, see \cref{app:CONTACT} for more details. 

The case of the contact terms for the amplitude in~\cref{Amp_Phi-bar{Phi}-phi-h} is simpler: it can be shown that the first contact term not vanishing after taking the classical limit comes at quadratic spin order.
Thus, $\M^{ \rm cl, (0)}_{C}\big [1_{\Phi}, 2_{\bar{\Phi}}, 3_\phi, 4_h^s \big ]=\M^{ \rm cl, (1)}_{C}\big [1_{\Phi}, 2_{\bar{\Phi}}, 3_\phi, 4_h^s \big ] = 0$. 
See \cref{sec:non-conformal} for a discussion on the contact terms at the quadratic order. 

In \cref{sec:Comparison} we discuss the approach commonly used to match these contact terms to specific solutions in a gravitational theory where black holes exhibit scalar hair. Apart from the approach studied there, another interesting avenue to perform this matching would be to use a modified version of Black Hole Perturbation Theory (BHPT), studying the scattering of a scalar wave on a black hole, similarly to what was done in~Ref.~\cite{Bautista:2022wjf} for Kerr BHs in GR. Indeed there has been some recent progress on alternative methods to extract quasi-normal modes in the context of scalar-tensor theories instead of relying on the standard BHPT in GR~\cite{Glampedakis:2019dqh,Silva:2019scu,Wagle:2021tam,Bryant:2021xdh,Li:2022pcy,Hussain:2022ins,Li:2023ulk,Wagle:2023fwl}. 

\subsection{Residues of 5-point amplitudes} 
\label{sec:residues}

In the  Kosower, Maybee, O'Connell (KMOC) framework, the input to calculate leading order classical radiation in collisions of compact objects is the classical limit of the on-shell 5-point amplitude
$\M[ (p_1+w_1)_{\Phi_1}  (p_2+w_2) _{\Phi_2} (-p_1) _{\bar \Phi_1}  (-p_2) _{\bar \Phi_2} (-k)_r]$~\cite{Cristofoli:2021vyo}, where 
$w_1+w_2= k$, and  $r$ denotes a massless quantum of radiation emitted with momentum $k$.
Within the PM expansion in gravitational scattering, this framework relies on a hierarchy of scales $\ell_c \ll \ell_{\rm Pl} \ll R_s \ll b$, 
where $\ell_c \sim 1/m$ is the Compton radius  and $\ell_{\rm Pl} \sim 1/\mpl$ is the Planck length.
As pointed out in Refs.~\cite{DeAngelis:2023lvf,Brunello:2024ibk}, what is actually sufficient is the knowledge of certain kinematic singularities (poles and cuts) of the amplitude. 
Moreover, at tree level, one only needs the residues of the poles at $w_{1,2}^2 \to 0$, which are readily computable from lower-point amplitudes using the unitarity methods. For our calculation of the waveforms we will need $R \equiv \text{Res}_{w_2^2 \to 0} \M^{\rm cl}[ (p_1+w_1) _{\Phi_1}  (p_2+w_2)_{\Phi_2} (-p_1)_{\bar \Phi_1}  (-p_2)_{\bar \Phi_2} (-k)_r]$,
as the residue at $w_1^2 \to 0$ is simplify related to $R$ by exchanging $(1 \leftrightarrow 2)$. 

First, we discuss the residues of the amplitude describing emission of a real scalar in matter scattering. 
The residues are related by unitarity to the three- and four-point amplitudes calculated earlier, in particular 
\begin{align}
R_X  = -  \sum_{i = h,\phi}
\M_X^{\rm cl}[ (p_1+w_1)_{\Phi_1} (-p_1) _{\bar \Phi_1}  (-k)_\phi (w_2)_i ] 
\M[(p_2+w_2)_{\Phi_2} (-p_2)_{\bar \Phi_2} (-w_2)_i ]     
, \end{align}
with $X=\{ U,C \}$.
For the presentation purpose we have split 
$R = R_{U} + R_{C}$, 
where the former factor descends from the unitary part of the 4-point amplitude, and the latter from the contact part. Furthermore, we will expand the latter in powers of spin: 
$R_C = R_C^{(0)}+ R_C^{(1)}  + \dots $.  
We find 
\begin{align}
    R_{U} =  &  \frac{2}{\mpl^3} \Bigg \{  \frac{(k w_2) [ (p_1 p_2)^2 - \frac{1}{2} m_1^2 m_2^2]+ m_2^2 (p_1 k)^2  -2 (p_1 p_2) (p_1 k) (p_2 k)  }{(p_1 k)^2} c_1 m_1^2 \cosh (w_2 a_2) 
    \nnl
	+ &  i  \frac{ c_1 m_1^2 (p_1 p_2)}{(p_1 k)^2} p_1^\mu p_2^\nu k^\rho w_2^\sigma \varepsilon_{\mu \nu \rho \sigma}  \sinh(w_2 a_2) -2 \frac{ c_2 m_2^2 (p_1 k)^2 \cosh (w_1 a_1)+ c_1 m_1^2 (p_2 k)^2 \cosh (w_2 a_2)}{w_1^2} 
    \nnl
	+ &\frac{2 i \varepsilon_{\mu \nu \rho \sigma}  p_1^\mu k^\nu w_2^\rho }{w_1^2} \Big [ a_1^\sigma (p_1 k) \frac{c_2 m_2^2 \sinh (w_1 a_1)}{w_1 a_1} + p_2^\sigma \frac{c_1 m_1^2 (p_2 k)}{p_1 k}  \sinh(w_2 a_2) \Big ] \Bigg \} , \label{Scalar-Res1U}
\end{align}
\begin{align}
   R^{(0)}_{C} =
   \frac{C_1^{(0)} c_2 m_1^2 m_2^2}{\mpl^3 } 
   , \label{Scalar-Res1C0}
\end{align}
\begin{align}
     R^{(1)}_{C} = 
     - i \frac{ C_1^{(1)} c_2 m_1^2 m_2^2}{\mpl^3} (w_1 a_1)
     . \label{Scalar-Res1C1}
\end{align}

We remark that  it is not straightforward to construct the full 5-point amplitude starting from $R_{U}$ in~\cref{Scalar-Res1U}.  
The obstacle is the presence of the $w_1^2$ pole in $R_{U}$, whose residue in the present form is not $1 \leftrightarrow 2$ symmetric. 
Fortunately, the residues are sufficient to calculate the waveforms, and we do not attempt to reconstruct the complete 5-point amplitudes. 
On the other hand, the contact parts  in~\cref{Scalar-Res1C0} and \cref{Scalar-Res1C1} do not contain any $w_1^2$ poles, and the corresponding contributions to the amplitude can be trivially reconstructed, 
$\M_5 \supset R_C^{(n)}/w_2^2 + (1 \leftrightarrow 2)$. 

Moving to graviton emission in matter scattering,  we denote 
 $R^s  \equiv\text{Res}_{w_2^2 \to 0}  \M^{s, \rm cl} [ (p_1+w_1) _{\Phi_1}  (p_2+w_2) _{\Phi_2} (-p_1) _{\bar \Phi_1}  (-p_2) _{\bar \Phi_2} (-k)_h^s]$.
It can be computed using 
\begin{align}
R_X^s   = -  \sum_{i = h,\phi}
\M_X^{\rm cl}[ (p_1+w_1)_{\Phi_1} (-p_1) _{\bar \Phi_1}  (-k)_h^s (w_2)_i ] 
\M[(p_2+w_2)_{\Phi_2} (-p_2)_{\bar \Phi_2} (-w_2)_i ]     
.  \end{align}
The  graviton-cut contribution is the same as in GR and, using the same integration methods and a four-vector representation of the amplitude,  it was first computed in Ref.~\cite{DeAngelis:2023lvf}.
Corrections beyond GR appear via the scalar cut. 
Up to the linear order in spin, there are no contact terms of the 4-point amplitudes to consider there  (they start at the quadratic order, see~\cref{Amp_Phi-bar{Phi}-phi-h_Contact}). 
For an incoming negative helicity graviton we find the non-GR pieces:
\begin{align} 
\Delta R_U^- = - \frac{c_1 c_2 m_1^2 m_2^2}{4 \mpl^3(p_1 k)^2 } \frac{\langle 5| w_2  p_1| 5\rangle^2} { w_1^2} ,
\label{Grav-Res2U}
\end{align}
where $| 5\rangle$ and $| 5]$ are the spinors corresponding to the momentum of the graviton.

%

\section{Waveforms} 
\label{sec:waveforms} 

The spectral waveform at tree level (leading PM order) is related  to a certain integral of the 5-point amplitude~\cite{Cristofoli:2021vyo}: 
\begin{align}
 \label{eq:WF_spectral-of-M} 
f_r(\omega) =  {1 \over 64 \pi^3 m_1 m_2} \int d \mu    
\M^{\rm cl}[ (p_1+w_1) _{\Phi_1}  (p_2+w_2) _{\Phi_2} (-p_1) _{\bar \Phi_1}  (-p_2) _{\bar \Phi_2} (-k)_r]|_{k=\omega n}
. \end{align}
In the present context, $r= \phi$ (scalar waveform) or $r=h$ (gravitational strain/waveform). 
The integration measure is 
$d\mu \equiv  
\delta^4(w_1 + w_2 - k)     \Pi_{i=1,2} [e^{i b_i w_i}  d^4 w_i \delta (u_i w_i)  ] $, where $u_i = p_i/m_i$ are the classical velocities of the scattering objects, and $b_i$ are their impact parameters. 
Finally, $\omega$ is the radiation frequency, and 
$n^\mu = (1,\boldsymbol{n})$ with 
$\boldsymbol{n} = (\sin \theta \cos \phi, \sin \theta \sin \phi, \cos \theta )$ a unit vector parametrizing the direction of emitted radiation. 

Following the steps in Ref.~\cite{DeAngelis:2023lvf} one can rewrite the waveform in terms of the residues of the 5-point amplitude: 
 \begin{align}
 \label{eq:WF_spectral}
 f_r(\omega) =  &  
 - { 1 \over  64 \pi^2 m_1 m_2  \sqrt{\gamma^2- 1} }  
\int_{-\infty}^\infty {d z  e^{i b_1 k+ i z (\hat u_1 k) b} \over \sqrt {z^2+1} }  
{1 \over 2} \bigg \{  
\nnl & 
R\big (w_2 \! \to  \!  (\hat u_1 k) \big [ 
\gamma \hat u_2  \! - \!  \hat u_1   \!+ \!  z \tilde b  \! +\!  i  \sqrt{z^2 \!+ \! 1} \tilde v   \big ]  \big )   
+ R \big (w_2 \! \to  \! (\hat u_1 k) \big [ 
\gamma \hat u_2 \! - \! \hat u_1  
\! + \!  z \tilde b  \! - \! i    \sqrt{z^2\! + \! 1} \tilde v   \big ]  \big ) 
 \bigg \} 
\nnl &  
  +  (1 \leftrightarrow 2)  
,   \end{align}
where $\hat u_n^\mu  \equiv u_n^\mu/\sqrt{\gamma^2-1}$, 
$b \equiv \sqrt{-(b_1 - b_2)^2}$, 
$\gamma \equiv u_1 u_2$, 
$\tilde b^\mu \equiv (b_1^\mu - b_2^\mu)/b$, 
$v^\mu \equiv \epsilon^{\mu\nu\alpha \beta} u_{1\nu} u_{2\alpha} \tilde b_\beta$, 
and $\tilde v^\mu \equiv v^\mu/\sqrt{-v^2}$.
See \cref{app:IntRes} for more details. 
The waveform in the time domain is the inverse Fourier transform of the spectral waveform 
\begin{align}
W_r =& 
\int_{-\infty}^\infty {d \omega \over 2\pi} e^{- i \omega t} f_r(\omega) 
,  \end{align}
where $t$ is the retarded time. 
We will expand the waveforms in powers of spin: 
$f_r = f_r^{(0)}+f_r^{(1)} + \dots$, 
and correspondingly 
$W_r = W_r^{(0)}+W_r^{(1)} + \dots$. 

\subsection{Scalar Waveforms}

We start with presenting the waveforms for scalar emission. 
Inserting \cref{Scalar-Res1U,Scalar-Res1C0,Scalar-Res1C1} into the master formula \cref{eq:WF_spectral}, expanding to linear order in  powers of spin, and transforming to the time domain we get 
  \begin{align}   
\label{eq:WF_Wphi0}
 W_\phi^{(0)}  =   &  
   - {  m_1 m_2  \over 64 \pi^2  \mpl^3  b   (\hat u_1 n)^2   \sqrt{\gamma^2- 1} \sqrt{ z^2 + 1}  }  
\re    \bigg \{ 
 {  2 (\gamma^2-1)  \big [  c_2 (\hat u_1 n)^2  +  c_1 (\hat u_2 n)^2    \big ]     
 \over
 \gamma  (\hat u_2 n) - (\hat u_1 n)  +  (\tilde b n)  z + i (\tilde v n) \sqrt{ z^2 + 1}    } 
         \nnl  &  
 +   c_1 {  ( 2  \gamma^2   - 1 ) z (\tilde b  n)  
-  (\hat u_1 n) 
- \gamma(2 \gamma^2 - 3 )  (\hat u_2 n)  
   \over   \gamma^2-1  } 
            + C_1^{(0)} c_2    (\hat u_1 n)   
\bigg \}|_{z=T_1}
 + (1 \leftrightarrow 2 )  
,     \end{align} 
   \begin{align}  \hspace{-2.5cm}  
   \label{eq:WF_Wphi1} 
 W_{\phi }^{(1)}  =   &   
  {m_1 m_2  \over 32  \pi^2  \mpl^3  b^2  (\hat u_1 n)^2  } 
{\partial \over \partial z } \Bigg ( \frac{1}{\sqrt{z^2+1}} \re  \bigg \{
 c_1  \big [  z (\tilde v n)   -  i  \sqrt{z^2+1}  (\tilde b n)   \big ]  
  \big [  - (\hat u_1 a_2)  +  z (\tilde b a_2) + i \sqrt{z^2+1} (\tilde v a_2)  \big ]   
 \nnl   \hspace{-2.6cm}  \times & 
   \bigg [   { \gamma  \over \gamma^2-1 }  
-     {    (\hat u_2 n)   \over 
 \gamma    (\hat u_2 n) -  (\hat u_1 n) + z (\tilde b n) + i\sqrt {z^2+1} (\tilde v n)   }  \bigg ] 
-        {  c_2   (\hat  u_1 n)    \over  
 \gamma    (\hat u_2 n) -  (\hat u_1 n) + z (\tilde b n) + i \sqrt {z^2+1} (\tilde v n)   } 
 \nnl   \hspace{-2.6cm}  \times &  
         \bigg [
\big [    i \sqrt{z^2+1}  (\tilde b n) -   z   (\tilde v n)  \big ] (\hat u_2 a_1) 
+  \big [   \gamma  (\tilde v n) +     i \sqrt{z^2+1}  \big ( \gamma (\hat u_1 n)  - \hat u_2 n \big )    \big ]  (\tilde b a_1)  
+  \big [   z    \big ( \hat u_2  n  - \gamma (\hat u_1 n) \big )   -  \gamma  (\tilde b n)    \big ] (\tilde v a_1)   
\bigg ]   
\nnl  \hspace{-2.6cm}  &  
  +  {   C_1^{(1)}  c_2    \over 2 \sqrt{\gamma^2- 1} }
\bigg [   (\hat u_1 n) \big [  \gamma (\hat u_2 a_1)  +  z (\tilde b a_1)   \big ]   -(a_1 n)    \bigg ] 
\bigg \} \Bigg ) |_{z=T_1}  + (1 \leftrightarrow 2)
,  \end{align} 
where 
 \begin{align} 
   T_i \equiv {t - b_i n \over   (\hat u_i n)  b }
   . \end{align}

The time-domain waveforms are fully analytic but lengthy. 
To gain more insight it is beneficial to approximate them in different regimes. 
First,  consider the large-time asymptotics,  $t \to \pm \infty$,  which corresponds to a large $z$ expansion of the waveforms.  
At the leading order in spin we find 
  \begin{align} 
\label{eq:WF_Wphi0larget}
 W_\phi^{(0)}  \to    &  
\mp  {m_1 m_2 c_1\over 64 \pi^2 \mpl^3 (\gamma^2-1)^{3/2} } \bigg \{ 
 (2 \gamma^2- 1  ) { (\tilde b n) \over (\hat u_1 n )^2 b} 
 + \bigg [  1  - (\gamma^2-1) C_2^{(0)} 
-   \gamma (2 \gamma^2-3) {( \hat u_2 n ) \over   (\hat u_1 n  )  }  \bigg ]  {1 \over t}
 \bigg \} 
 \nnl & 
  + (1 \leftrightarrow 2)  + \cO(t^{-2})
  .  \end{align} 
The asymptotics contains a constant term, related to the scalar memory effect, see \cref{sec:memory}. 
The leading time-dependent piece decays as $1/t$, consistently with the
   $f_\phi^{(0)} \sim \omega^0$ long-wavelength behavior of the spectral waveform. 
It controls the asymptotic form of the power emitted in scalar radiation. 
At the linear order in spin
  \begin{align} 
\label{eq:WF_Wphi1larget}
 W_\phi^{(1)}  \to    &  
  \pm {m_1 m_2 c_1\over 64 \pi^2 \mpl^3 (\gamma^2-1) } \bigg \{ 
2 \gamma  { (\tilde b n) (\tilde v a_2) + (\tilde v n) (\tilde b a_2)
 \over (\hat u_1 n )^2 b^2} 
+ \bigg [ \gamma \big [ (\tilde b n) (\tilde v a_2) + (\tilde v n) (\tilde b a_2)   \big ] 
  \nnl   &  
+ \sqrt{\gamma^2-1} C_2^{(1)} \big [ 
(\tilde b n) (\tilde b a_2) + (\tilde v n) (\tilde v a_2) 
+  { (u_1 n )  ( u_1 a_2 ) \over \gamma^2-1}\big ]
\bigg ] {1 \over t^2} 
\bigg \}+   (1 \leftrightarrow 2) + \cO(t^{-3})
.  \end{align}

We move to another limit of the waveforms: 
$\gamma  = {1 \over \sqrt{1-v^2}}  \to 1$. 
In this limit the two bodies scatter with non-relativistic relative velocity $v \ll 1$. 
This is the relevant limit for typical astrophysical objects, such as NSs,  BHs, etc, which we do not expect to have relativistic velocities. 
For concreteness, let us restrict to the rest frame of particle $2$:  
$u_2 = (1,\boldsymbol{0})$, 
$u_1 = (\gamma,\sqrt{\gamma^2-1} \boldsymbol{\hat v})$,
with $\boldsymbol{\hat v}^2 = 1$. 
We parametrize the impact parameters and spins as 
$b_1 = (0, b \boldsymbol{\hat b})$,
$a_1  =  L.(0, \boldsymbol{a_1})$, 
$a_2  =  (0, \boldsymbol{a_2})$, 
where 
$\boldsymbol{\hat b}^2= 1$, 
and $L$ the Lorentz boost from the rest frame of the particle $1$. 

At the leading order in spin expansion for $|t| \lesssim b/v$ 
  \begin{align}   
  \label{eq:WF_Wphi0nr}
 W_\phi^{(0)} = & 
 {m_1 m_2 (c_1-c_2) \over 64 \pi^2   \mpl^3 b  }
\bigg \{ 
- {(\boldsymbol{\hat v} \boldsymbol{\hat n}) \over v}
+    ( \boldsymbol{\hat b} \boldsymbol{\hat n}) {t \over  b}
\bigg \} + \cO(v) 
  .  \end{align}  
The first term in the curly bracket is time-independent, 
while the one with linear time dependence controls the power emitted during closest approach in non-relativistic scattering. 
Note that the leading order result does not depend on the Wilson coefficients $C_i^{(0)}$ of the contact terms (those enter only at $\cO(v^2)$). 
Note also that, at that order, the waveform is proportional to the difference of the scalar charges, and it vanishes for $c_1 = c_2$. 
This is a consequence of the $1 \leftrightarrow 2$ symmetry of the waveform  and of the fact that, at this PN order, there is no radiation  perpendicular to the $\boldsymbol{\hat v}-\boldsymbol{\hat b}$ plane.
Note that, at higher PN orders, the waveform does not vanish for  $c_1=c_2$.

At the subleading order in spin expansion the non-relativistic result is more complicated: 
 \begin{align}  
   \label{eq:WF_Wphi1nr}
W_\phi^{(1)} = & 
 {m_1 m_2  \over 32 \pi^2 \mpl^3 b^2  }  \bigg \{ 
 \big [ c_2 \boldsymbol{a_1}-c_1 \boldsymbol{a_2}\big ] \cdot \boldsymbol{\hat v} 
\big ( \boldsymbol{\hat v} \times \boldsymbol{\hat b} \cdot \boldsymbol{\hat n} \big ) 
+ {1 \over 2}\big [ c_1 C_2^{(1)}  \boldsymbol{a}_2 - c_2 C_1^{(1)}  \boldsymbol{a}_1 \big ]  \cdot \boldsymbol{\hat b}
\nnl & 
+ v \bigg [ \big [ c_1 \boldsymbol{a_2}  -  c_2 \boldsymbol{a_1}  \big ] 
\cdot \big [ 
2  \boldsymbol{\hat b}    \big ( \boldsymbol{\hat v} \times \boldsymbol{\hat b} \cdot \boldsymbol{\hat n} \big ) 
+ \boldsymbol{\hat v} \times \boldsymbol{\hat b}   \big ( \boldsymbol{\hat b} \boldsymbol{\hat n} \big )   
\big ] 
+  {1 \over 2} \big [ c_1 C_2^{(1)}  \boldsymbol{a}_2 - c_2 C_1^{(1)}  \boldsymbol{a}_1 \big ] \cdot \boldsymbol{\hat v} 
\bigg ] {t \over b}
\bigg \}  
\nnl &  
+ \cO(v^2) 
  .  \end{align}  
The spinning part is suppressed by one power of velocity compared to the zeroth order result in~\cref{eq:WF_Wphi0nr}.

\subsection{Gravitational Waveforms}

We move to gravitational radiation in our framework. 
For reference, we first quote the waveform in GR at the leading and subleading order in spin~\cite{Jakobsen:2021lvp,DeAngelis:2023lvf,Aoude:2023dui,Brandhuber:2023hhl}. 
As is customary, we calculate the waveform defined by \cref{eq:WF_spectral-of-M} with the 5-point amplitude describing emission of a negative helicity graviton (the analogous object with a positive helicity graviton carries the same physical information). 
At the zeroth order in spin we find the following compact representation of the time-domain waveform expressed in terms of helicity spinors:  
 \begin{align}  
 \label{eq:WF_Wh0GR}
W_{h}^{{\rm GR} (0)}   =  &    
   -  { m_1 m_2 
   \over 512 \pi^2   \mpl^3   b   (\hat u_1 n)^2   \sqrt{\gamma^2 -1}    } 
     {1 \over  \sqrt {z^2+1}} 
        {\cal R} \bigg \{   
      { \big (\mathcal{F}_1^-(z) \big )^2 + \big (\mathcal{F}_2^-(z) \big )^2    
                \over  
       \gamma    (\hat u_2 n) -  (\hat u_1 n) + z (\tilde b n) + i \sqrt {z^2+1} (\tilde v n)       } \bigg \}_{|z=T_1}
       \nnl    & 
    +  (1 \leftrightarrow 2) 
.   \end{align} 
Here ${\cal R}$ acts as the real part operation, $\re[x] = (x+x^*)/2$, except it is defined not to conjugate the helicity spinors and sigma matrices. 
At the linear order in spin
\begin{align}  \hspace{-2.6 cm}
 \label{eq:MW_Wh1GR}
W_h^{{\rm GR}(1)} = & 
\frac{i m_1 m_2}{512 \pi^2 \mpl^3\sqrt{\gamma^2-1} (\hat u_1 n)^3 b^2} \frac{\partial}{\partial z} \Bigg ( \frac{1}{\sqrt{z^2+1}} \mathcal{R} \bigg \{  \frac{1}{ \gamma (\hat u_2 n) - \hat u_1 n+z (\tilde b n) + i \sqrt{z^2+1}(\tilde v n)} 
\nnl  \hspace{-2.6 cm}  &
\times \bigg [(a_1 n) \big [  \big (\mathcal{F}_1^-(z) \big )^2 + \big (\mathcal{F}_2^-(z) \big )^2  \big ]  + (\hat u_1 n) (\gamma \hat u_2^\mu - \hat u_1^\mu+ z \tilde b^\mu + i \sqrt{z^2+1} \tilde v^\mu) (a_{1 \mu} + a_{2 \mu}) \big [ \big (\mathcal{F}_1^-(z) \big )^2  - \big (\mathcal{F}_2^-(z) \big )^2  \big ] \nnl \hspace{-2.6 cm} &
\qquad + 2 (\hat u_1 n) \mathcal{F}_1^- (z) \mathcal{F}_3^-(z) \bigg ]   \bigg \} 
\Bigg )_{|z=T_1}
          +  (1 \leftrightarrow 2)  
,  \end{align}
and
\begin{align}
    \mathcal{F}_1^-(z) &=\langle n| \big ( \hat u_1 \hat u_2+2\gamma z \hat u_1 \tilde b +z \tilde b \hat u_2 + 2 i \gamma \sqrt{z^2+1} \hat u_1 \tilde v + i \sqrt{z^2+1} \tilde v \hat u_2 \big) | n \rangle , \nnl
    \mathcal{F}_2^-(z) &=\langle n|\big ( \hat u_1 - z \tilde b - i \sqrt{z^2+1} \tilde v \big ) \hat u_2 | n \rangle , \nnl
    \mathcal{F}_3^-(z) &= \langle n| \big ( \hat u_2 - \gamma \hat u_1 + (\gamma^2-1) \hat u_1 \hat u_2 (z \tilde b + i \sqrt{z^2+1}\tilde v) \big) a_1| n \rangle \label{F_functions_definition} 
. \end{align}

We can also define the $\mathcal{F}^+_i(z)$ functions by replacing angle with square brackets. 
The positive helicity counterpart of the waveform can be obtained from the expression above by complex conjugation, and it does not carry any additional information.
In \cref{F_functions_definition} we have also used the notation $|n\rangle \equiv  \frac{| 5\rangle}{\omega}$, $|n] \equiv  \frac{| 5]}{\omega}$.

We discuss now the corrections to the GR waveform due to the presence of the scalar $\phi$ with monopole coupling to macroscopic objects. 
Due to the nature of the scalar coupling the correction only affects the zeroth order waveform. 
We find
 \begin{align}  
  \label{eq:WF_DeltaWh0}
\Delta W_h^{(0)}   =  &    
  -{ c_1 c_2 m_1 m_2    \over 512 \pi^2  \mpl^3   b (\hat u_1 n)^2  \sqrt{\gamma^2- 1}} 
 {1 \over \sqrt {z^2+1} } {\cal R} \bigg \{  
{ \langle n| \big ( \gamma \hat u_2   +  z \tilde b  + i \sqrt{z^2+1} \tilde v \big )\hat u_1  | n \rangle^2 
\over 
 \gamma    (\hat u_2 n) -  (\hat u_1 n) + z (\tilde b n) + i \sqrt {z^2+1} (\tilde v n)      }   \bigg \}|_{z=T_1} 
\nnl & 
   +  (1 \leftrightarrow 2) 
.   \end{align}  

For $t \to \pm \infty$, the asymptotic form of the waveform is obtained by  expanding the above at large $|z|$. 
We get 
 \begin{align}  
W_{h}^{(0)}   \to  &    
   \mp   { m_1 m_2 
   \over 512 \pi^2   \mpl^3   b   (\hat u_1 n)^2   \sqrt{\gamma^2 -1}    } 
        {\cal R} \bigg \{   
 \nnl &
 { \big \langle n| \hat u_2 \big(\tilde b+ i \tilde v \big)|n \rangle^2  
      + \langle n | \big ( 2  \gamma   \hat u_1    - \hat  u_2 \big ) \big ( \tilde b + i  \tilde v \big ) | n \rangle^2 
  + c_1 c_2 \langle n| \hat u_1 \big ( \tilde b + i \tilde v \big )| n \rangle^2   
                \over  
     (\tilde b n) + i  (\tilde v n)        } 
     \bigg \}
       \nnl   &
    +  (1 \leftrightarrow 2) + \cO(1/t) 
.   \end{align} 
The scalar contributions affect in a non-trivial way the memory effect. 

In the rest frame of particle $2$ and with non-relativistic relative velocity we find the approximation 
\begin{align}
 \label{eq:WF_Wh0nr}
\partial_t W_h^{(0)} = & 
- e^{-2 i \phi}  {m_1 m_2  \over 16 \pi^2 \mpl^3 b^2  }  
\bigg [ 1  + {c_1 c_2 \over 2 } \bigg ]
 \bigg \{ 
(\boldsymbol{\hat v} \boldsymbol{\hat n})(\boldsymbol{\hat b} \boldsymbol{\hat n}) 
+ i (\boldsymbol{\hat v} \times \boldsymbol{\hat b}) \cdot \boldsymbol{\hat n} \bigg \} v  
+ \cO(v^2)
. \end{align}
It can be remarked that, at the leading order in the velocity expansion, the effect of the scalar charge on the spinless gravitational  waveform is a simple change of normalization.
In particular, the gravitational waveform remains suppressed by one power of velocity, as in GR. 
The consequence is that gravitational wave emission at the leading PN order has the same angular distribution as in GR, possibly up to spin effects. 

\subsection{Emitted power}
\label{EmittedPower}

\begin{figure}
    \centering
    \includegraphics[width=0.8\linewidth]{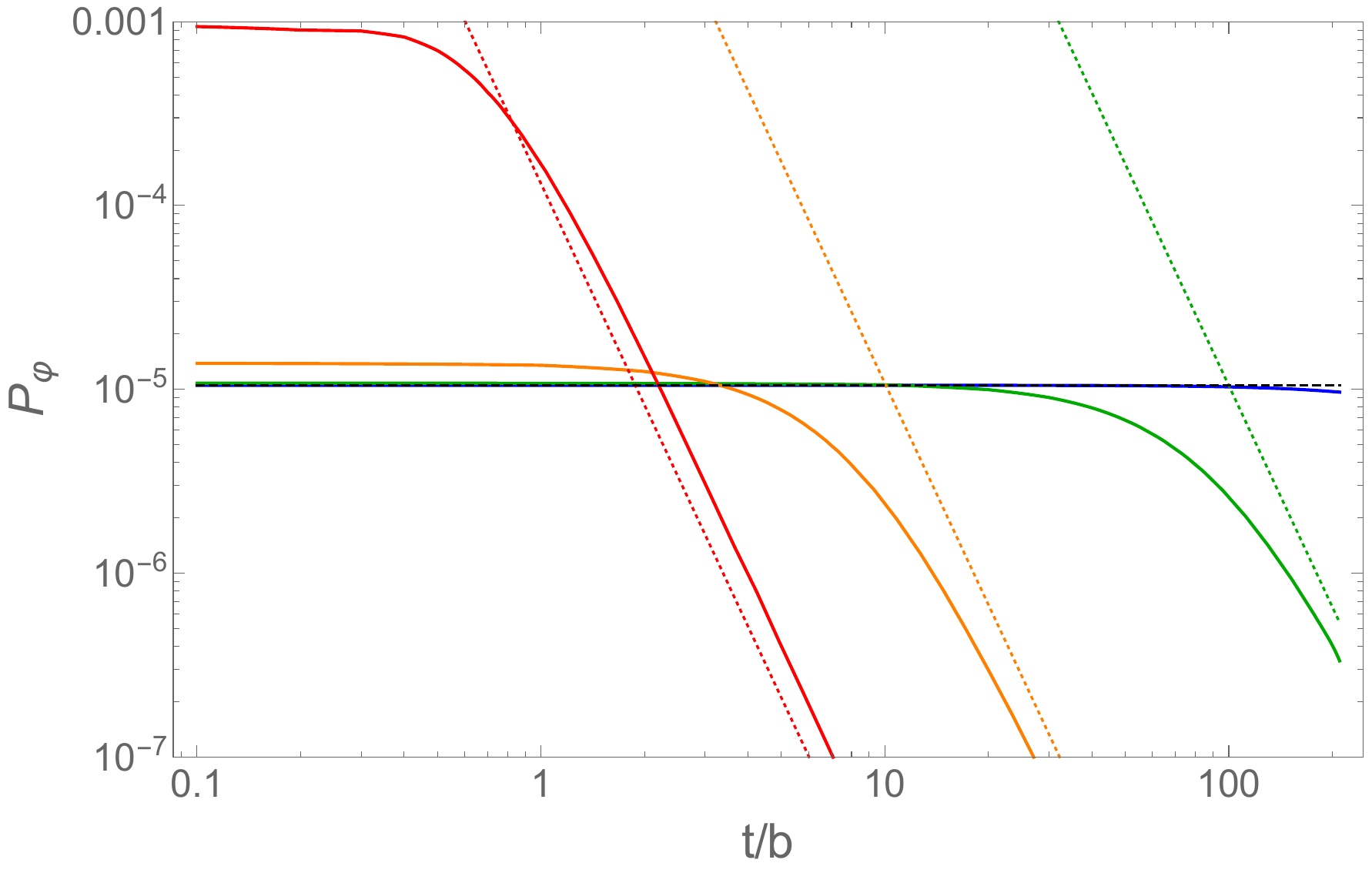}
    \caption{Power emitted in scalar radiation in the rest frame of the object $2$ in units of
    ${c_1^2 m_1^2 m_2^2 \over b^4 \mpl^6}$ at the zeroth order in spin calculated using the waveform in~\cref{eq:WF_Wphi0} for $c_2 = C_1^{(0)}= C_2^{(0)} =0$.  
    We show results for the velocity of particle $1$ fixed as $v=0.9$ (red), 
    $v=0.1$ (orange), $v=10^{-2}$ (green), and $v=10^{-3}$ (blue).
    The dotted lines correspond to power calculated using the large $t$ approximation of the waveform in~\cref{eq:WF_Wphi0larget}, and the dashed line uses the non-relativistic approximation in~\cref{eq:WF_Wphi0nr}. 
    }
    \label{fig:Power-scalar}
\end{figure}

\begin{figure}
    \centering
    \includegraphics[width=0.49\linewidth]{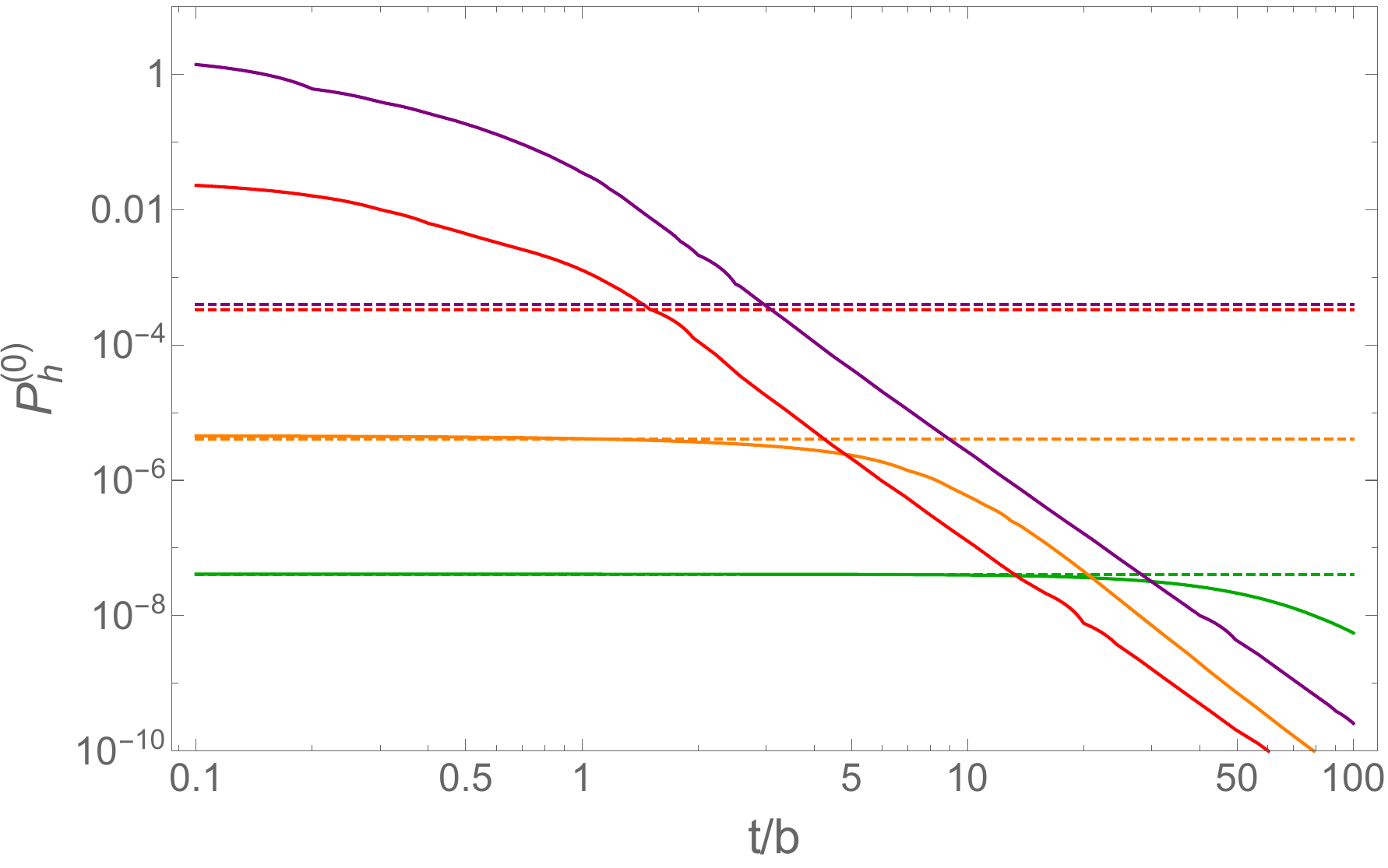}
        \includegraphics[width=0.49\linewidth]{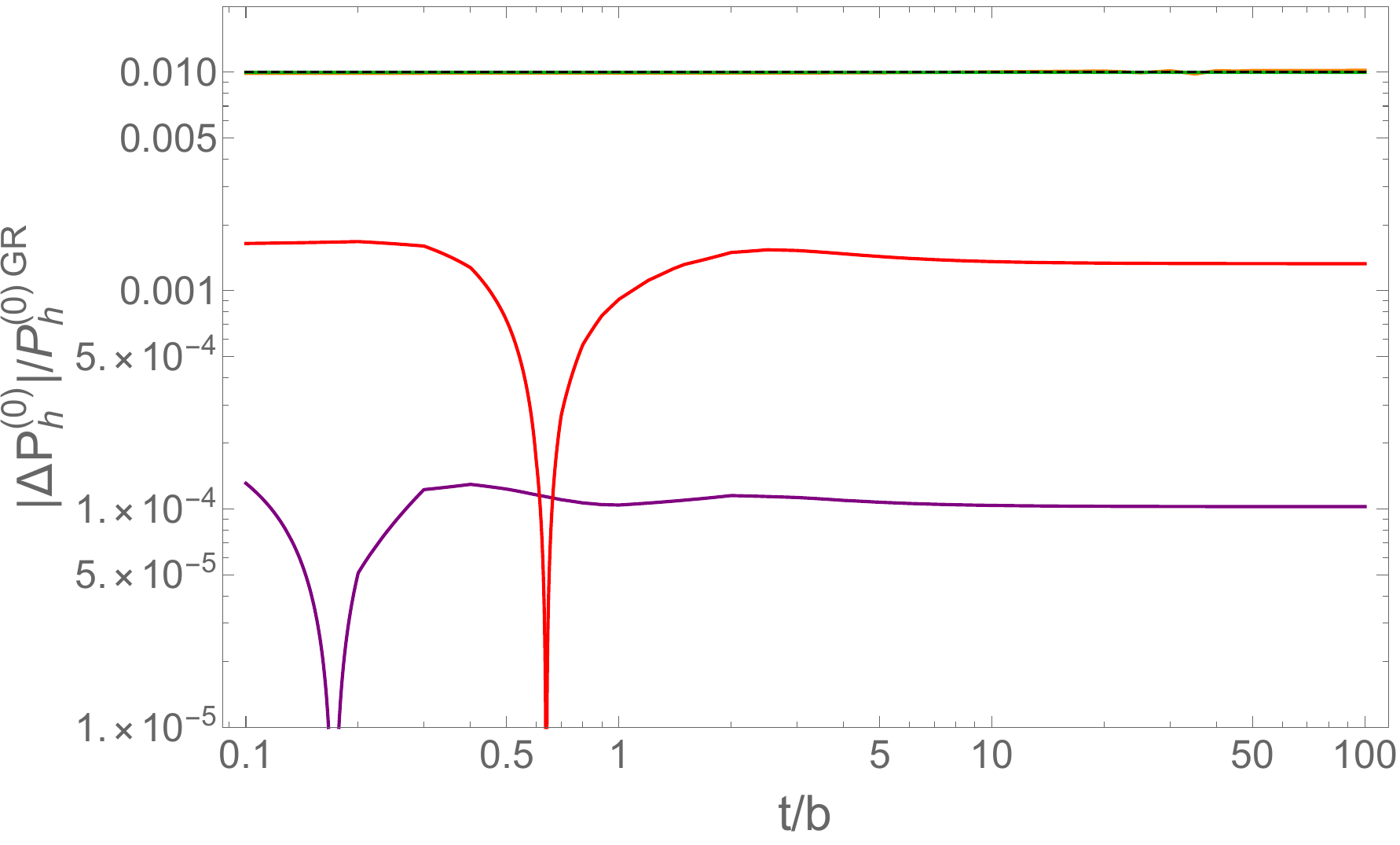}
    \caption{Left: power emitted in gravitational radiation in the rest frame of the object $2$ in units of
    ${m_1^2 m_2^2 \over b^4 \mpl^6}$ at the zeroth order in spin calculated using the waveform in~\cref{eq:WF_Wh0GR,eq:WF_DeltaWh0} for $c_1 = 1$, $c_2 = 0.01$. 
    We show results for the velocity of particle $1$ fixed as $v=0.99$ (purple), $v=0.9$ (red), 
    $v=0.1$ (orange), and $v=10^{-2}$ (green). 
    The dashed lines use the non-relativistic approximation in~\cref{eq:WF_P0nr}. 
    Left: the same for the relative difference between gravitational radiation in the scalar-tensor framework and in GR. 
    The dips are due to $\Delta P_h^{(0)}$ changing sign from negative to positive. The green and orange lines practically overlap with the non-relativistic approximation (black dashed). 
    }
    \label{fig:Power-gravity}
\end{figure}

We have all the elements to calculate the power emitted  during collision of two compact objects with scalar charges. 
The power emitted in gravitational and scalar radiation is related in a simple manner to the respective waveforms: 
 \begin{align}  
  \label{eq:WF_EmittedPower}    
{d P_h \over d \Omega} =  2   |\partial_t W_h |^2  
, \nnl 
{d P_\phi \over d \Omega} = (\partial_t W_\phi)^2  
   .   \end{align}    
Integrating over the angles we obtain the total emitted power as a function of time.  
We will expand the result in  powers of spin: 
$P_X = P_X^{(0)} +  P_X^{(1)} + \dots$. 
Up to any given order it is straightforward to perform the integration numerically using the full tree-level waveforms; 
in particular up to linear order in spin we need the waveforms in~\cref{eq:WF_Wphi0,eq:WF_Wphi1,eq:WF_Wh0GR,eq:WF_DeltaWh0,eq:MW_Wh1GR}. 
On the other hand, in the non-relativistic regime one can perform the integrals analytically.\footnote{It is also possible to derive a fully analytic PM result for the total radiated energy, by using the approach in  Ref.~\cite{Herrmann:2021lqe}. }
Indeed, using \cref{eq:WF_Wphi0nr,eq:WF_Wh0nr} one obtains 
\begin{align} 
\label{eq:WF_P0nr}
P_h^{(0)}   =  & {m_1^2  m_2^2 \over 80 \pi^3 \mpl^6 b^4 }
( 1 + c_1 c_2 )  v^2 
+ \cO(v^3) 
, \nnl 
P_\phi^{(0)}   =  &
 {m_1^2 m_2^2  \over 3072 \pi^3 \mpl^6 b^4 }(c_1-c_2)^2    + \cO(v)
. \end{align} 
These approximations are compared to the full numerical results in~\cref{fig:Power-scalar,fig:Power-gravity}. Note that, in this limit,  the result does not depend on the leading order contact terms $C_i^{(0)}$, even though the complete waveform does. 
This means that radiation at low velocities is universal at the leading order, in the sense that the details of the interaction Lagrangian or concrete solutions in  theories that exhibit scalar hair for each compact object are condensed into a single parameter: the scalar monopole charge. We also recover the result~\cite{Yagi:2011xp} that scalar radiation from collisions of compact objects with monopole scalar charge enters at  ${-1\rm PN}$ order compared to the gravitational radiation: 
\begin{align} 
{P_\phi^{(0)} \over P_h^{(0)}  } \approx {5 (c_1-c_2)^2  \over 192 v^2 }
. \end{align} 
Therefore, for typical astrophysical objects, scalar radiation is the dominant mechanism of anomalous energy loss, which is also one of the intuitive reasons that theories exhibiting monopole scalar hair are so phenomenologically relevant. 
We obtain this result for hyperbolic scattering, but it matches the conclusions obtained by other means for bounds systems~\cite{Yagi:2011xp,Shiralilou:2021mfl}. 
We continue our results to bounds systems by assuming the Kepler law 
${b^3 \over T^2} ={ G (m_1 + m_2) \over 4 \pi^2}$,
$T= \frac{m_1 +m_2}{4 \mpl^2 v^3}$,
which, for quasi-circular orbits,  allows us to trade the impact parameter $b$ for velocity: 
$b^{-1} \to  { 8 \pi \mpl^2 v^2 \over m_1 + m_2}$ \footnote{It is also possible to use the map in Refs.~\cite{Kalin:2019rwq, Kalin:2019inp, Cho:2021arx} to analytically continue results for the total radiated momentum from hyperbolic to elliptic orbits, see for example Eq. (11) of~\cite{Herrmann:2021lqe}}. 
Then, \cref{eq:WF_P0nr} becomes 
\begin{align} 
\label{eq:WF_P0-kepler}
P_h^{(0)}   =  & {256 \pi m_1^2  m_2^2 \mpl^2 (1 + c_1 c_2)  \over 5 (m_1 + m_2)^4 }  v^{10}
+ \cO(v^{11}) 
, \nnl 
P_\phi^{(0)}   =  &
 {4 \pi m_1^2 m_2^2  \mpl^2 (c_1-c_2)^2
 \over 3 (m_1 + m_2)^4  }  v^{8}   + \cO(v^9)
. \end{align} 
The first one is the well-known formula for gravitational wave emission in GR, see~\cite{Misner:1973prb, Blanchet:1989cu}, rescaled by the  factor $1+ c_1 c_2$. 
The second agrees with the leading PN and PM result in Ref.~\cite{Shiralilou:2021mfl} (see their eq.~61) with the matching  $\alpha_n^0 = c_n/\sqrt 2$ (the overall $\frac{1}{\sqrt{2}}$ factor amounts to a different normalization of the scalar field in that reference). 
It also agrees with Ref.~\cite{Yagi:2011xp} (see their eq.~133) with the identification 
$q_n = G \mpl m_n c_n = {c_n m_n \over 8 \pi \mpl}$.\footnote{Ref.~\cite{Yagi:2011xp} work with the units where $G=1$. 
In this map, we inserted one power of $G$ so as to make $q_n$ dimensionless in our system of units.}

It is also interesting to study the emitted power at the subleading PN order. 
In the center-of-mass frame of the collision the scalar power at the zeroth spin order in the PN expansion can be written 
$P_\phi^{(0)}  + \Delta_{\rm PN} P_\phi^{(0)} + \cO(v^3)$,  
where $P_\phi^{(0)}$ is the same as the one in \cref{eq:WF_P0nr} and 
  \begin{align}  
   \label{eq:CC_Delta2P0nrCM}
 \Delta_{\rm PN} P_\phi^{(0)} =  &
 {m_1^2 m_2^2 \over \pi^3 \mpl^6 b^4 } 
 \bigg  [ 
 {(c_1 + c_2)^2  \over 3840 } 
 +  \big ( 6 - \eta  \big )  { (c_1 - c_2)^2    \over 7680 } 
  +   {(c_1 + c_2) (c_1 - c_2)  \over 3840 } {m_1 - m_2 \over m_1+m_2} 
  \nnl & 
    +   {(c_1 - c_2) \big ( m_2    c_2       C_1^{(0)}        - m_1  c_1  C_2^{(0)}   \big ) 
     \over 1536 ( m_1 + m_2 ) } 
     -  {(c_1-c_2)^2  \over 1536} {t^2 \over b^2}  
 \bigg ] v^2 
  ,   \end{align}  
where $\eta \equiv {m_1 m_2 \over (m_1 + m_2)^2}$.
The $t$-independent terms above match the corresponding bound-orbit expression for the emitted power in Ref.~\cite{Shiralilou:2021mfl}\footnote{%
Effects of higher PM origin, including $\dot r$ pieces, as well as $\cO(\hat \alpha c_n)$ terms are ignored in this comparison.}  with the additional identification $C_n^{(0)} = - \beta_n^0$. 
The $t^2$ term in \cref{eq:CC_Delta2P0nrCM} eventually leads to a breakdown on our PN expansion for $|t| \gtrsim b/v$, which marks a transition to the large $|t|$ asymptotic behavior. 
It is not clear to us what is the counterpart of this term on the bound-orbit side, if any.

Moving to the next order in spin, the correction to the  power emitted in scalar radiation at the linear order in spin can be calculated from the waveforms via   
$ P^{(1)} = 2 \int d \Omega \partial_t W_\phi^{(0)}  \partial_t W_\phi^{(1)}$.
In the non-relativistic limit this can be approximated by 
  \begin{align}  
 P_\phi^{(1)}=  &
 {m_1^2 m_2^2 (c_1-c_2) \over 768 \pi^3 \mpl^6 b^5 } 
( \boldsymbol{\hat v} \times \boldsymbol{\hat b} ) 
\cdot \big [  c_1 \boldsymbol{a_2}  - c_2 \boldsymbol{a_1} \big ]   v 
  .  \end{align}  
This is non-zero only when at least one of the spins has a component orthogonal to the collision plane. 
The integrated radiated power is independent of the contact terms $C_i^{(1)}$, because their effect in~\cref{eq:WF_Wphi1nr} does not interfere with the leading order waveform in  \cref{eq:WF_Wphi0nr}. 
However, the angular distribution of radiated scalar power is affected by the contact terms at the linear order in spin.
Note that the leading spin correction is suppressed by one power of velocity and one power of the impact parameter. 
For quasi-circular orbits this translates into $v^3$ suppression relative to  $P_\phi^{(0)}$ in~\cref{eq:WF_P0-kepler}. 

\subsection{Memory effect}
\label{sec:memory}

We now aim to use the waveforms computed in the previous section to address the memory effect for both gravitational and scalar radiation. 
There has been recent interest into the potential observability of the memory effect~\cite{Strominger:2014pwa} and amplitudes methods have been already used to compute the memory effect in GR~\cite{Herderschee:2023fxh,Georgoudis:2023lgf,Aoude:2023dui,Brandhuber:2023hhl}. The waveforms computed in the presence of the conformal coupling do lead to a non-zero scalar memory effect and to modifications to the gravitational memory effect. 
We comment that in the setting of  Ref.~\cite{Falkowski:2024bgb} it was observed that there was no memory effect, which is  related to the shift symmetry of the interactions. However, in the presence of scalar hair that symmetry is not manifest, and  it is this distinction that leads to non-zero memory effects.

We start with the scalar memory effect - see \cite{Garfinkle:2017fre,Bernard:2022noq,DiVecchia:2022owy,Falkowski:2024bgb} 
for previous discussion of this effect in various contexts. 
Following the presentation in Ref.~\cite{Falkowski:2024bgb}, we define it as:
\begin{align}
    \Delta \phi = W_\phi (t = \infty) - W_\phi ( t = - \infty) .
\end{align}

We restrict our attention to the leading PM (and $c_n$) order memory effect and we make an expansion in the spin vector:
\begin{align}
    \Delta \phi = \sum_{n=0}^{\infty} \Delta^{(n)} \phi .
\end{align}
Taking the result for the scalar waveform at zeroth spin order \cref{eq:WF_Wphi0} we get the memory effect for the scalar up to linear spin order:
\begin{align}
    \Delta^{(0)} \phi = - \frac{ c_1 m_1 m_2 (2\gamma^2-1) }
    {32 \pi^2  (\gamma^2-1)^{3/2} b \mpl^3 } \frac{(\tilde b n)}{(\hat u_1 n)^2}  
    + (1 \leftrightarrow 2)  
    \label{ScalarMemory0}
, \end{align}
\begin{align}
	\Delta^{(1)} \phi = \frac{c_1 m_1 m_2 \gamma}{16 \pi^2 (\gamma^2-1) b^2 \mpl^3 }  \frac{ (\tilde b n) (\tilde v a_2) +  (\tilde v n) (\tilde b a_2)}{(\hat u_1 n)^2} + (1 \leftrightarrow 2) 
    \label{ScalarMemory1} 
. \end{align}

As expected on general grounds, contact terms do not contribute to the memory effect~\cite{Brandhuber:2023hhl}. 
Note that we could also relate the scalar memory effect to a soft theorem for scalars in this context~\cite{Falkowski:2024bgb}, but this is beyond the scope of our discussion.

Now we move to the gravitational memory effect. 
Our definition  is:
\begin{align}
    \Delta h^- = 
    W_h (t = \infty) - W_h ( t = - \infty),
\end{align}
where the superscript  marks that the waveforms are calculated for an incoming graviton of negative helicity. 
Notice that our convention for the waveforms is different from the literature: the gravitational strain commonly defined in GR  can be computed by multiplying our results with an overall $\kappa \equiv \frac{2}{\mpl}$ factor.
The memory effect gets a new contribution only at zeroth spin order:
\begin{align}
	\Delta h^{(0)}_{-}  = 
    - \frac{c_1 c_2 m_1 m_2}{256 \pi^2 (\hat u_1 n)^2 \mpl^3 \sqrt{\gamma^2-1} b[(\tilde b n)^2 + (\tilde v n)^2]}  
    & \bigg \{ (\tilde b n ) \Big [ \langle n | \tilde b \hat u_1 |n \rangle^2  - \langle n | \tilde v \hat u_1 | n \rangle^2  \Big ] \nnl
	& + 2 (\tilde v n) \langle n|\tilde b \hat u_1 |n \rangle \langle n | \tilde v \hat u_1 | n \rangle   \bigg \} + (1 \leftrightarrow 2).
\end{align}
The memory effect also in this case can be connected to the soft theorem and Weinberg's soft factor. 

The memory effects are related to what are referred as the \emph{static} contributions of the waveform. It is not clear whether these could be mapped from the bound to the scattering case~\cite{Adamo:2024oxy}. For this reason, we do not attempt to make a direct comparison of the above memory effects with other results in the literature for bound orbits, as in e.g.~Ref.~\cite{Bernard:2022noq}.

\subsection{Comparison to the  skeletonization approach} 
\label{sec:Comparison}
It is interesting to examine the relationship between our approach with the skeletonization procedure often used in the GR context. 
This was first introduced in Ref.~\cite{1975ApJ...196L..59E} and further expanded later in Refs.~\cite{Damour:1992we,Damour:1995kt} to account for the presence of a scalar field in multi-scalar tensor theories when studying compact objects.
In this approach, the gravitational action in the presence of matter in the Einstein frame is $S=S_{\rm vac} [g_{\mu \nu}, \phi] + S_{\rm m}[\Phi_n, \mathcal{A}^2(\phi)g_{\mu \nu}]$, with
\begin{align}
    S_{\rm m}[\Phi_n, \mathcal{A}^2(\phi)g_{\mu \nu}] = - \sum_A \int m_A(\phi) \sqrt{-g_{\mu \nu} dx_A^\mu  dx_A^\nu } ,
\end{align}
where $A$ runs from $1,...,n$, 
$n$ is the number of matter particles we consider, $x_A^\mu $ parametrizes the worldline of particle $A$ and we use the conventions of~\cite{Shiralilou:2021mfl} for this discussion. 
The constant GR mass is replaced by a function $m_A(\phi)$ depending on the values of the scalar field on the worldline. Note that this action does not contain derivatives of the scalar field $D_{\mu_1, ... , \mu_n} \phi$, which have been shown to be related to finite size-effects~\cite{Damour:1998jk}. The mass function can be parametrized in terms of
\begin{align}
    \alpha_A = \frac{d \ln m_A(\phi)}{d \phi} \qquad \qquad , \qquad \qquad \beta_A = \frac{d \alpha_A (\phi)}{d\phi} ,
\end{align}
by expanding them as
\begin{align}
    m_A(\phi) = m_A^0 \Big [ 1 + \alpha_A^0 \delta \phi + \frac{1}{2} \big ( (\alpha_A^0)^2 + \beta_A^0 \big) \delta \phi^2 + \mathcal{O}(\delta \phi^3) \Big] , \label{Mass_function}
\end{align}
where $\delta \phi = \phi - \phi_0$ and $\phi_0$ is a background scalar field value. The superscript "0" means that the respective quantity is evaluated at $\phi = \phi_0$. The coupling $\alpha_A^0$, often called in the literature "sensitivity" or "scalar charge", measures the coupling of the scalar field to the "skeletonized" body $A$. By selecting a specific theory that specifies $S_{\rm vac}$, one can then proceed to compute observables from a worldline formalism using the skeletonized action.

We now discuss the relation of these approaches to our results, focusing on the case of SGB theories. 
Firstly, the "scalar charge" $\alpha_A^0$ is mapped to our conformal coupling $c_A$ (e.g. at the level of the observables, such as the energy flux), which is also mapped to the scalar charges $q_A$ defined in Ref.~\cite{Yagi:2011xp}, see \cref{EmittedPower}. 
Moreover, in this formalism, the $\beta_A^0$ parameter encodes the same information as our contact term $C_A^{(0)}$ - and thus one can obtain a map between the two as well. 
This effect was not considered in Ref.~\cite{Yagi:2011xp}, but was taken into account by later works on the SGB theories~\cite{Julie:2019sab,Shiralilou:2021mfl}. 
This contact term does not affect the scalar energy flux at the leading PN order (it contributes to higher PN orders as was shown in Ref.~\cite{Julie:2019sab}), which is in perfect agreement with our expression in the second line of \cref{eq:WF_P0nr}. 

Once the theory is specified, one can then use a matching procedure to derive analytic expressions for $\alpha_A^0$, $\beta_A^0$ - and also the higher order ones we neglected in~\cref{Mass_function}. As long as the compact objects have secondary scalar hair, these parameters can be written in terms of the coupling of the theory and the BH parameters (mass, charge and angular momentum). This is done by studying BH solutions in scalar-tensor theories and their thermodynamic properties and matching to the effective worldline description of the scalar dependent mass functions presented above.. This has already been done up to fourth order in the coupling constant in SGB~\cite{Julie:2019sab}. 
In particular, given that 
$S_{\rm vac} [g_{\mu \nu}, \phi] = 
\frac{c^3}{16 \pi G} \int_M d^4x \sqrt{-g} \Big [ R -2 (\nabla \phi)^2 + \alpha f(\phi) R_{GB}^2 \Big ] $, at first order in the coupling the scalar charge reads $\alpha^0_A = - \frac{\alpha f'(\phi_0)}{2m_A^2}$ \footnote{Note that, according to~\cite{Julie:2019sab} we should replace the mass parameter with the irreducible mass $m_{\text{irr}, A}$ here, but this makes no difference for static and spherically symmetric BHs, as the ones studied in~\cite{Shiralilou:2021mfl}.}.
It would be interesting to use a similar procedure to study a broader spectrum of modified gravity theories and also include finite-size and spin effects. 
This will allow for the explicit expressions of the couplings related to spinning effects, like the $C_A^{(1)}$ contact term of our approach.\footnote{A first study of spin effects appeared very recently in the literature~\cite{Almeida:2024cqz}. Unfortunately this reference does not quote the emitted power, therefore we can not easily match their results to our approach.} 

\section{Beyond conformal coupling}
\label{sec:non-conformal}

In the previous section we studied classical radiation in theories with a massless scalar conformally coupled to spinning matter. 
We found that, in this setting, the leading PN contribution matches the behavior due to the monopole scalar charge in SGB scalar-tensor theories.
In this section we investigate how to parametrize departures from conformal coupling  within the on-shell amplitude framework, and discuss some possible interpretations of such deformations.

\subsection{Contact term domination}
\label{sec:CTD}

It is worth to first consider a particular limit of the scalar waveforms studied in~\cref{sec:waveforms} where the contact terms dominate over the monopole scalar charges. 
Formally, for the waveforms in~\cref{eq:WF_Wphi0,eq:WF_Wphi1} we apply the limit  $C_n^{(k)} \to \infty$, $c_n \to 0$, with 
$C_n^{(k)} c_n$ finite.  
The non-relativistic limit of the waveforms obtained this way reads 
 \begin{align}  
\label{eq:WF_Wphi1nr-contact}
\partial_t W_\phi^{(0)} = & \cO(v^2) 
, \nnl 
\partial_t W_\phi^{(1)} = & 
 {m_1 m_2  \over 64 \pi^2 \mpl^3 b^3  } \Big (
 \big [ c_1 C_2^{(1)}  \boldsymbol{a}_2 - c_2 C_1^{(1)}  \boldsymbol{a}_1 \big ] \cdot \boldsymbol{\hat v}  \Big ) v
+ \cO(v^2) 
  .  \end{align} 
In this limit the waveform at the zeroth order in spin in suppressed. 
As a consequence, the leading PN behavior of the emitted power is quadratic in spin: 
 \begin{align}  
 \label{eq:BCC_Phi2-C1}
P_\phi^{(2)}  = & 
{m_1^2 m_2^2  \over 1024 \pi^3 \mpl^6 b^6  } 
 \bigg ( \big [ c_1 C_2^{(1)}  \boldsymbol{a}_2 - c_2 C_1^{(1)}  \boldsymbol{a}_1 \big ] \cdot \boldsymbol{\hat v}  \bigg )^2  v^2  
.   \end{align} 
Qualitatively, this resembles the formula for power emitted in the DCS scalar-tensor theory, with $C_n^{(1)} \boldsymbol{a_n}$ playing the role of dipole scalar charges. 
Connecting to bound state systems by trading the impact factor for velocity using the Kepler law, one concludes $P_\phi^{(2)} \sim v^{14}$, in agreement with the behavior of the DCS result in Ref.~\cite{Yagi:2011xp}.
However, the latter result is of the form 
$P_\phi^{(2)} \sim  
(\boldsymbol{a} \boldsymbol{v})^2 + \# \boldsymbol{a}^2$, 
and \cref{eq:BCC_Phi2-C1} does not reproduce the second term, in particular it vanishes for spins orthogonal to relative velocity. 
We speculate that the lack of quantitative matching may be due to a non-trivial boundary-to-bound continuation. 

We may also interrogate whether other contact terms could affect the power emitted at the quadratic order in spin. 
We already exhausted contact terms at the zeroth and linear order in spin, but we have not yet studied those at the quadratic order. 
In principle, these could interfere with the zeroth order waveform in~\cref{eq:WF_Wphi0} in emitted power, and dominate the $\cO(c^2)$ terms in a suitable limit of large Wilson coefficients. 
For the amplitude describing scalar-graviton scattering off a matter particle, the unique contact term at the quadratic order can be parametrized as: 
\begin{align} 
\label{Amp_Phi-bar{Phi}-phi-h_Contact}
\M_C^{(2)} \big [1_{\Phi_n} 2_{\bar\Phi_n} 3_\phi 4_h^- \big ]  = & { {\hat D_n^{(2)} } \over {\mpl^2}}  \langle 4| p_1 a_n | 4\rangle^2  
 ,     \nnl 
 \M_C^{(2)} \big [1_{\Phi_n} 2_{\bar\Phi_n} 3_\phi 4_h^+ \big ]  = & { {\hat{D}_n^{(2)*}}  \over {\mpl^2}}  [ 4| p_1 a_n | 4]^2,
 \end{align} 
where $\hat D_n^{(2)} = D_n^{(2)} + i \tilde D_n^{(2)}$ is a dimensionless Wilson coefficient.
In the non-relativistic limit, 
the parity-even coefficient  $D_n^{(2)}$ contributes to the  scalar waveform as
\begin{align}  
\partial_t W_{\phi}^{(2)} \supset &  
 {3 m_1 m_2 \over 4 \pi^2 \mpl^3 b^4  }  \bigg \{ 
 D_1^{(2)} (\boldsymbol{a_1} \boldsymbol{\hat v}) (\boldsymbol{a_1} \boldsymbol{\hat b})
+ D_2^{(2)}  (\boldsymbol{a_2} \boldsymbol{\hat v}) (\boldsymbol{a_2} \boldsymbol{\hat b})   \big  ] 
\bigg \}v  + \cO(v^2) 
    .  \end{align}
On the other hand, the parity-odd coefficient $\tilde D_n^{(2)}$ contributes to the scalar waveform as
\begin{align}  
\partial_t W_{\phi}^{(2)} \supset & 
-  {3 m_1 m_2  \over 4 \pi^2 \mpl^3 b^4  }  \bigg \{ 
 \tilde D_1^{(2)} ( \boldsymbol{\hat v}  \boldsymbol{a_1}) 
 ( \boldsymbol{\hat v}\times  \boldsymbol{\hat b}  \cdot \boldsymbol{a_1})  
 -  \tilde D_2^{(2)} ( \boldsymbol{\hat v}  \boldsymbol{a_2}) 
 ( \boldsymbol{\hat v}\times  \boldsymbol{\hat b}  \cdot \boldsymbol{a_2})   
 \bigg \} v^2 + \cO(v^3) 
    .  \end{align}  
In both cases, this affects the angular distribution of the emitted scalar power, but does not interfere with \cref{eq:WF_Wphi0nr} to contribute to the total power.     

We may also consider contact terms in the amplitude describing a scalar scattering off a matter particle,
The zeroth and first order contact terms in that amplitude were written down in~\cref{Amp_Phi-bar{Phi}-phi-phi_Contact0,Amp_Phi-bar{Phi}-phi-phi_Contact1}. 
At the quadratic level we have 
   \begin{align} 
   \label{eq:BCC_MphiphiC2}
\M_{C}^{\rm cl} \big [1_{\Phi_n} 2_{\bar \Phi_n} 3_\phi 4_\phi  \big ]  = &  C_n^{(2)} {m_n^2 \over \mpl^2}    (q a_n)^2    
 .   \end{align} 
This contributes to the scalar waveform as
\begin{align}  
\partial_t W_{\phi}^{(2)} \supset & 
 v {3 m_1 m_2  \over 32 \pi^2 \mpl^3 b^4  }  
 \bigg \{ C_1^{(2)} c_2 ( \boldsymbol{\hat v}  \boldsymbol{a_1}) ( \boldsymbol{\hat b}  \boldsymbol{a_1}) 
 + C_2^{(2)} c_1 ( \boldsymbol{\hat v}  \boldsymbol{a_2}) ( \boldsymbol{\hat b}  \boldsymbol{a_2}) \bigg \}  
 .   \end{align}  
This does not interfere with  the zeroth order waveform in~\cref{eq:WF_Wphi0nr}. 
We conclude that, when the contact term contribution dominates the 5-point scalar emission amplitude, \cref{eq:BCC_Phi2-C1} is a robust prediction for the emitted power at the leading PN order.
Moreover, for $C_n^{(0)} = 0$, in this limit there is no scalar radiation at all at the zeroth and first order in spin, much as in the DCS framework\footnote{This is a consequence of the fact that in DCS it is only spinning BHs that possess a dipole scalar charge~\cite{Yagi:2012vf}}. Notice that, since this consists of only a contact term's contribution, it does not produce any memory effect at leading order.

\subsection{Spin-dependent scalar couplings}
\label{sec:Spin_dependent_couplings}

We have argued that, in the classical limit, conformal coupling of matter to the massless scalar $\phi$ translates into a spin-independent 3-point amplitude in~\cref{Amp_phi-Phi-Bar{Phi}}. 
But, obviously, this is not the most  general coupling between the scalar and matter. 
In this subsection we discuss possible classical deformations of that amplitude, which lead to a spin-dependent scalar-matter coupling. 
The most general form of that 3-point amplitude in the classical limit reads 
\begin{align}
\label{eq:BCC_MphiPhiPhi}
 \M^{\text{cl}} \big [1_{\Phi_n}, 2_{\bar \Phi_n} , 3_\phi\big ] = & 
     -  \sum_{k = 0}^\infty g_n^{(k)}  \frac{m_n^2}{\mpl}  (i p_3 a_n)^k ,
\end{align}
where $g_n^{(k)}$ are real couplings by crossing symmetry, 
and  $g_n^{(0)}$ is identified with $c_n$ in~\cref{Amp_phi-Phi-Bar{Phi}}. 
The interaction with $n$-th power of spin in~\cref{eq:BCC_MphiPhiPhi} can be obtained from quantum-level amplitudes similar to the ones in~\cref{Amp_3_bos} but proportional to  
$ \big (  \langle \bm 2 \bm 1 \rangle  - [ \bm 2 \bm 1] \big )^n$.  
It is reasonable to speculate that $g_n^{(k)}$ parametrize the internal structure of the compact object represented by $\Phi_n$.
The effect of these couplings on the 5-point classical emission amplitudes  can be easily read off from the results in~\cref{sec:residues} by replacing  $c_n \to  \sum_k g_n^{(k)} (-i w_n a_n)^k$. 

As an example, consider the situation when $g_n^{(1)} \neq 0$ while the remaining $g_n^{(k)}$ vanish. 
The scalar waveform then takes the form 
   \begin{align} 
 W_\phi^{(1)}  =   &     
 {  m_1 m_2  \over  32  \pi^2 (\hat u_1 n)^2   b^2 \mpl^3   \sqrt{\gamma^2- 1} }  
{d \over d z} \bigg (  {1 \over \sqrt {z^2+1} }  
 \re  \bigg \{  
  g_1^{(1)}   \bigg   [ { (a_1 n ) \over     (\hat u_1 n) } 
  - \gamma (\hat u_2 a_1) -  z (\tilde b a_1) - i  \sqrt{z^2+1} (\tilde v a_1) 
\bigg ]  
 \nnl  \times & 
 \bigg [ 2  \gamma        (\hat u_2 n) -     (\hat u_1 n) 
-  { \gamma^2   -  {1 \over 2 } \over \gamma^2 - 1}  
 \big [ \gamma (\hat u_2  n) - (\hat u_1 n)   +  z( \tilde b n) + i  \sqrt{z^2+1} (\tilde v n)        \big ] 
 \bigg ]   
\nnl  - &
   {\gamma^2-1   \over    \gamma (\hat u_2  n) - (\hat u_1 n)   +  z( \tilde b n) + i  \sqrt{z^2+1} (\tilde v n)          }  
 \bigg [    
 g_1^{(1)}     (\hat u_2 n )^2  \big [ { (a_1 n) \over (\hat u_1 n) } 
  -  \gamma (\hat u_2  a_1)  - z (\tilde b a_1) - i  \sqrt{z^2+1}( \tilde v a_1)  \big ]
  \nnl  - &  
    g_2^{(1)}  (\hat u_1 n )^2   \big [ (\hat u_1  a_2) -  z (\tilde b a_2) - i  \sqrt{z^2+1} (\tilde v a_2)  \big ]
  \bigg ] 
+ \frac{  g_1^{(1)}  C_1^{(0)}  }{2} (\hat u_1 n) 
\big [ (\hat u_1 a_2) - z (\tilde b a_2) \big] \bigg \}
   \bigg )_{|z=T_1}
     \nnl   &  +  (1 \leftrightarrow 2) 
,   \end{align}
where we also included the contact term in~\cref{Amp_Phi-bar{Phi}-phi-phi_Contact0}. 
In the non-relativistic limit, the power emitted in scalar radiation starts at the second order in spin is
\begin{align}
\label{eq:BCC_scalarP2nr}
P_\phi^{(2)} = {m_1^2 m_2^2 \over 1024 \pi^3 b^6 \mpl^6} \bigg \{ 
\bigg (   \big [  g_1^{(1)} \boldsymbol{a_1}  + g_2^{(1)} \boldsymbol{a_2} \big]^2  \cdot \boldsymbol{\hat b} \bigg ) ^2 
+ {1 \over 3} \big [  g_1^{(1)} \boldsymbol{a_1}  + g_2^{(1)} \boldsymbol{a_2} \big]^2 
\bigg \}
. \end{align}
This enters at order $\cO(v^0)$, much as in the case of conformal coupling in~\cref{eq:WF_P0nr}. 
But the fact that that the leading effect in this set-up appears at the second order in spin implies additional suppression by two powers of $S/(b m)$.
Continuing \cref{eq:BCC_scalarP2nr} to quasi-circular bound states leads to $P_\phi^{(2)} \sim v^{12}$,  
thus suppressed by two PN orders relative to the conformal coupling case in~\cref{eq:WF_P0-kepler}, but enhanced by one PN order relative to the contact term domination scenario in~\cref{sec:CTD}.

The correction to the gravitational waveform reads 
\begin{align}  
    \Delta W_h^{(2) - } = &
     \frac{g_1^{(1)} g_2^{(1)} m_1 m_2}{512 \pi^2 \mpl^3 \sqrt{\gamma^2-1}(\hat u_1 n)^2 b^3} \frac{\partial^2}{\partial z^2} \Bigg \{ 
    \frac{1}{\sqrt{z^2+1}} \mathcal{R} \bigg [ 
    \frac{ \langle n| \big ( \gamma \hat u_2+ z \tilde b+i \sqrt{z^2+1} \tilde v \big ) \hat u_1 | n \rangle^2 }
    {\gamma (\hat u_2 n)-(\hat u_1 n)  + z (\tilde b n ) + i \sqrt{z^2+1} (\tilde v n) } 
    \nnl \times &
    \big [(\hat u_1 a_2) - z (\tilde b a_2) - i \sqrt{z^2+1} (\tilde v a_2) \big ] \big [ \frac{(n a_1)}{(\hat u_1 n)} - \gamma (\hat u_2 a_1) - z (\tilde b a_1) - i \sqrt{z^2+1} (\tilde v a_1) \big ] \bigg ] \Bigg \}_{|z=T_1}
         \nnl   &  +  (1 \leftrightarrow 2)  
. \end{align}
With this waveform one can show that, much as for the conformal coupling, the power radiated in gravitational waves is subleading by one PN order ($\cO(v^2)$) compared to the scalar power.
The memory effect can easily be derived from the above formula by taking the large $z$ limit of  the expression in the curly brackets.

\section{Conclusions} 
\label{sec:conclusions}

In this paper we studied scalar-tensor theories of gravity using the on-shell amplitude methods. 
Our main focus was on modeling classical collisions of compact objects that carry monopole scalar charges.
An effective description of compact objects in such a setting is provided by massive spinning matter particles coupled to gravity and with a conformal coupling to the massless scalar. 
We provided a set of on-shell 3-point amplitudes that define this framework.  
These basic building blocks are related to higher-point amplitudes via the unitarity cuts. 
With the help of the unitary factorization techniques, the 3-point amplitudes allowed us to calculate the residues of 5-point amplitudes describing emission of radiation quanta in a collision of two matter particles. 
There exists an ambiguity in this calculation due to the possibility of adding contact terms in the intermediate 4-point amplitudes. 
We parametrize it by explicitly listing the relevant contact terms up to linear spin order, and give a prescription for constructing contact terms at any spin order. 

Using the KMOC formalism, the residues of the 5-point amplitudes can be connected to the waveforms for emission of scalar and gravitational radiation in hyperbolic collision of compact objects. 
We calculated these waveforms at the leading PM order and up to linear order in spin.  
The power emitted in hyperbolic collisions was studied and, after a continuation to bound systems, compared  to the known results obtained in the classical theory framework. 
At the leading PN order we found the matching between our results and the classical ones wherever analytic results were available. 

An interesting limiting case of this framework is when the pieces deriving from the contact terms dominate the intermediate 4-point amplitudes. 
This leads to a qualitatively different pattern of radiation, which reproduces features of radiation from compact objects with scalar dipole charges. 
Such objects are found for example in the DCS scalar-tensor theory, where spinning BHs do not have monopole scalar charges but instead carry spin-dependent dipole charges. 
We also studied qualitatively the consequences of departing from the paradigm of conformal coupling  between the massless scalar and matter.

There are several advantages of our approach.
The calculation uses only gauge-invariant ingredients with clear physical interpretation. 
Explicit compact object solutions in scalar-tensor theories are not needed to calculate the emitted power (though they will be useful to match the relevant couplings and contact terms in theories that exhibit secondary hair, like SGB and DCS). 
The waveforms are valid for any relative velocity of the scattering objects, in particular they can also be used in the relativistic regime $v \sim 1$.  
Furthermore,  they can be systematically improved to include higher powers of spin, or higher PM effects (in the amplitude formalism corresponding to loop corrections to the 5-point amplitude  as well as some cut contributions). 
One downside is that the waveforms are calculated for hyperbolic scattering, which are much less frequently observed by experiment. 
As of now, there is no systematic procedure to  continue the results to bound-state systems, although recent approaches show great promise~\cite{Adamo:2024oxy}. 
Nevertheless, many observables related to the binary dynamics - including the energy emitted - can be analytically continued from the scattering to the bound case~\cite{Kalin:2019rwq,Kalin:2019inp,Cho:2021arx}.

The scalar coupling to spinning compact objects within our framework is designed to model binaries in scalar-tensor theories, but the applications should not be limited to that. 
We believe that a similar effective treatment can be used to address other types of new physics, related to scalar particles. 
Some ideas include models involving scalar fields coupled to nucleons, such as the QCD axion, altering the description of NSs~\cite{Balkin:2023xtr}, soliton and boson stars~\cite{DelGrosso:2023trq,Boskovic:2021nfs,Bezares:2022obu} or even boson clouds surrounding BHs~\cite{Baumann:2022pkl,Boskovic:2024fga,Takahashi:2024fyq}. In these models a more careful treatment may be needed to account for massive scalar fields, their possible self-interactions or the equation of state in the case of NSs. 
Still, as we have established in this work, the on-shell amplitude approach together with the KMOC formalism severely limit the set of possible couplings that can contribute to classical phenomena.

Another interesting application of our model is in the context of standard GR. There has been recent work on the comparison between the PM and self-force expansions~\cite{Barack:2023oqp} by employing a scalar charge toy model. This provides a simpler environment to crosscheck calculations and evaluate the accuracy of numerical methods, which can be severely complicated for the gravity case~\cite{Quinn:2000wa,Gralla:2021qaf,Barack:2022pde}. 
One can verify that in fact the model presented in Ref.~\cite{Barack:2023oqp} is a limiting case of our framework, corresponding to ignoring the spin effects ($a_n \to 0$) and setting one of the two conformal couplings and the contact terms to zero ($c_2 \to 0$, $C_n^{(i)} \to 0$ with $c_1 \neq 0$).\footnote{Specifically, translating to the conventions of that paper the identification should be $c_1 \to \frac{4 \sqrt{\pi} \mpl Q}{m_1}$.}  The present work provides a consistent and systematic method to go beyond that framework and incorporate additional effects, such as the ones related to spinning observables or incorporation of scalar charges for both compact objects, which could be very useful for further comparisons of the PM and self-force results.

The are other interesting avenues to pursue in the future, apart from the ones mentioned above. 
Since we discuss spin, a direct extension of our work would be to include a matching calculation of our couplings and contact terms beyond the zeroth order in the spin, e.g. along the lines of \cite{Almeida:2024cqz}. 
Apart from that, it is  intriguing  to better understand the description of higher "multipole scalar charges", given our discussion for the dipole charge in DCS in \cref{sec:CTD}. 
Obviously, a better understanding of the physical correspondence of the deviations from the conformal couplings that we studied in \cref{sec:Spin_dependent_couplings} is also in order. 
Other types of deviations from the conformal couplings can also be implemented in our approach and their potential detectability can be studied. Finally, the calculations presented in this work can be straightforwardly extended to study other classical observables as well as higher PM orders within our setup.

\section*{Acknowledgements}

We would like to thank Christos Charmousis and Karim Noui for useful discussions. 
AF has received funding from the Agence Nationale de la Recherche (ANR)  grant ANR-19-CE31-0012 (project MORA) and from the European Union’s Horizon 2020 research and innovation programme  the Marie Skłodowska-Curie grant agreement No 860881-HIDDeN. 

\appendix

\newpage
\section{Conventions} 
\label{app:CON}

In this appendix be briefly summarize our conventions. 
We  work with the mostly minus metric $\eta_{\mu \nu} = (1,-1,-1,-1)$, 
and the natural units $\hbar = c = 1$.
The sign convention for the totally anti-symmetric Levi-Civita tensor $\epsilon^{\mu\nu\rho\alpha}$ is 
$\epsilon^{0123} = 1$. 
The 2-index antisymmetric tensor is defined as $\epsilon^{\alpha \beta}$ with $\epsilon^{12}=1$.  
The Lorentz-invariant integration measure is 
$d\Phi_k \equiv {d^4 k \over (2\pi)^4 } \delta(k^2-m^2)\theta(k^0)$. 
For 3-vectors we use the bold notation: 
$\boldsymbol{x} \equiv \vec{x}$.  
We use the all-incoming convention for our on-shell amplitudes unless stated otherwise.
In the context of GR, the Christoffel connection is
$ \Gamma^\mu_{\ \nu \rho} =   
{1 \over 2}  g^{\mu \alpha} \left (
\partial_\rho g_{\alpha \nu} +  \partial_\nu g_{\alpha \rho} - \partial_\alpha g_{\nu \rho} \right ) 
$, 
the Riemann tensor is 
$R^\alpha_{\ \mu \nu \beta}  =   
 \partial_\nu \Gamma^\alpha_{\ \mu \beta} -  \partial_\beta \Gamma^\alpha_{\ \mu \nu}  
 +  \Gamma^\rho_{\ \mu \beta}  \Gamma^\alpha_{\ \rho \nu}
 - \Gamma^\rho_{\ \mu \nu}  \Gamma^\alpha_{\ \rho \beta}  
$, 
and the Ricci tensor is 
$R_{\mu \nu}  =  R^\alpha_{\ \mu \nu \alpha}$. The Gauss-Bonnet and the Chern-Simons are respectively defined as $R_{\rm GB}^2 \equiv 
R^{\mu \nu  \rho \sigma}  R_{\mu \nu  \rho \sigma}
- 4R^{\mu \nu}  R_{\mu \nu} + R^2$ and $R_{\rm CS}^2  \equiv R^{\mu \nu  \rho \sigma}  \tilde R_{\mu \nu  \rho \sigma}$, with $\tilde R_{\mu \nu}{}^{ \alpha \beta } \equiv  
{1 \over 2 } \epsilon^{ \alpha \beta \rho \sigma}  R_{\mu \nu  \rho \sigma} $.
We use the all-incoming convention for our on-shell amplitudes, unless otherwise noted.  
For helicity spinors, we use the conventions of Ref.~\cite{Dreiner:2008tw}, plus the usual shorthand notation: 
$\lambda_p^\alpha \equiv |p\rangle$, 
$\tilde \lambda_p^{\dot \alpha} \equiv |p]$  
for massless 2-component spinors, 
$\alpha=1,2$, 
and 
$\chi_k^{\alpha\, J} \equiv |\bm k\rangle^J$, 
$\tilde \chi_k^{\dot \alpha\, J} \equiv |\bm k \rangle^J$ for massive spinors, with the little group index $J=1,2$ encoding polarization and being usually suppressed in our formulas. 
In our conventions,
$\langle k l \rangle \equiv
\lambda_k^\alpha \lambda_{l\alpha}$,
$[ k l ]\equiv
\tilde \lambda_{k\dot \alpha} \tilde \lambda_l^{\dot \alpha}$, 
$\langle k | p | l ] \equiv
\lambda_k^\alpha [p \sigma]_{\alpha \dot \beta} \tilde \lambda_l^{\dot \beta}$, 
$[k | p | l \rangle \equiv
\tilde \lambda_{k\dot \alpha} [p \bar \sigma]^{\dot \alpha \beta} \lambda_{l\beta}$,
$\langle k | pq | l \rangle \equiv
\lambda_k^\alpha [p \sigma]_{\alpha \dot \beta}[q \bar \sigma]^{\dot \beta \beta} \lambda_{ l \beta}$,
$[k | p q | l ]\equiv
\tilde \lambda_{k\dot \alpha} [p \bar \sigma]^{\dot \alpha \beta} [q \sigma]_{\beta \dot \beta} \tilde \lambda^{\dot \beta}_{l}$
and similarly for more four-vector insertions, with 
$\sigma^\mu \equiv (1,\vec{\sigma})$,
$\bar \sigma^\mu \equiv (1,-\vec{\sigma})$.

\section{Integrals} 
\label{app:IntRes}

In this appendix we make a quick overview of our method to compute the leading order integrals which give the time domain waveform. Our approach follows closely the lines of Ref.~\cite{DeAngelis:2023lvf}. 
We depart from the method presented in Appendix~A of Ref.~\cite{Falkowski:2024bgb} which applies to the same type of integrals when the complete five-point amplitudes  are readily available. Here, we tackle the same problem in the cases where only the residue is available.

The general type of integral we encounter is
\begin{align}
    I = \int d\mu \M_5
, \end{align}
where the integral measure is defined as:
\begin{align}
d \mu \equiv &  \delta^4(w_1 + w_2 - k)     \Pi_{i=1,2} [e^{i b_i w_i}  d^4 w_i \delta (u_i w_i)  ] ,
, \end{align}
and $\M_5 \equiv \M[(p_1 + w_1)_{\Phi_1} (p_2 + w_2)_{\Phi_2} 
(-p_1)_{\bar \Phi_1} (-p_2)_{\bar \Phi_2} (-k)_r]$ defines the five-point amplitude describing the emission of a massless quantum $r$. The spectral waveform $f$ is related to the integral via
$f_r = {I \over 64 \pi^3 m_1 m_2}$.

We deal with the integrals using the parametrization that first appeared in Ref.~\cite{Cristofoli:2021vyo}:
\begin{align}
    w_2^{\mu} = z_1 u_1^{\mu} + z_2 u_2^{\mu} + z_\upsilon \tilde{\upsilon}^\mu + z_b \tilde{b}^\mu, 
\end{align}
where
\begin{align}
     \centering \upsilon^\mu \equiv \varepsilon^{\mu \alpha \beta \rho} u_{1 \alpha} u_{2 \beta} \tilde{b}_\rho \ , \qquad \qquad \tilde{\upsilon}^\mu = \frac{\upsilon^\mu}{\sqrt{-\upsilon^2}} ,
\end{align}

\begin{align}
    \centering b \equiv \sqrt{-(b_1-b_2)^2} \ , \qquad \qquad \tilde{b}^\mu = \frac{b_1^\mu - b_2^\mu}{b} 
,\end{align}
and $u_n = p_n/m_n$. 
We also define $ \hat u_i \equiv {u_i \over \sqrt{\gamma^2-1} }$. We then perform first the $z_1, z_2$ integrals which set $z_1 = - \frac{(\hat{u}_1 k)}{\sqrt{\gamma^2-1}}$ and $z_2 = \frac{\gamma (\hat{u}_1 k)}{\sqrt{\gamma^2-1}}$ \footnote{We refer the interested readers to Appendix A of~\cite{Falkowski:2024bgb} for details.}. At this stage we have but to integrate over the two remaining variables:

\begin{align}
    I = \frac{e^{ib_1 k }}{\sqrt{\gamma^2-1}} \int dz_v d z_b e^{i b z_b } \M_5 |_{z_1, z_2} , \label{Int_zvzb}
\end{align}
where
\begin{align}
    w_2^\mu |_{z_1, z_2} &=
    (\hat u_1 k) [\gamma \hat u_2^\mu - \hat u_1^\mu] + z_b \tilde b^\mu + z_v \tilde v^\mu \qquad \quad &,
    , \nnl  w_2^2 |_{z_1, z_2} &= 
    - [ (\hat u_1 k) + z_v^2 + z_b^2 ]
    , \nnl
    w_1^\mu |_{z_1, z_2} &=
    (\hat u_2 k) [ \gamma \hat u_1^\mu - \hat u_2^\mu] - [z_b + (\tilde b k)] \tilde b^\mu - [ z_v + (\tilde v k)] \tilde v^\mu  
 , \nnl
 w_1^2 |_{z_1, z_2} &= - [ (\hat u_2 k) + z_v^2 + z_b^2 ] 
, \end{align}
and $|_{z_1, z_2}$ implies that we have set the values of $z_1, z_2$ as described before.

At this point one can make a choice whether to first integrate \cref{Int_zvzb} over $z_v$ or over $z_b$. This will lead to two different but equivalent representations of the integral, which we call the $z_b$ and $z_v$ representations, respectively.

The $z_b$ representation amounts to the choice in Ref.~\cite{DeAngelis:2023lvf}. 
In this case, one integrates over $z_v$ by first computing the residue in the upper half-plane, denoted $I_+$, and then in the lower half-plane, denoted $I_-$ with the poles being at $z_v = \pm i \sqrt{z_b^2 + (\hat u_1 k)^2}$ and $z_v = \pm i \sqrt{(z_b + \tilde b k)^2 + (\hat u_2 k)^2} - \tilde{v}k$. The final result is the difference between the two $I = \frac{I_+ - I_-}{2}$, so as to cancel the pole at infinity. 
After some simple manipulations one can put the result in a symmetric form:

\begin{align}
    I &= - \frac{\pi}{2 \sqrt{\gamma^2-1}} \int_{-\infty}^\infty 
    \frac{dz}{\sqrt{z^2+1}}  \bigg \{ 
     e^{i b_1 k + i z (\hat u_1 k) b} \bigg [  
     {\rm Res}_{w_2^2\to 0 } \M_5{}_{|z \to R_2}  
     +  {\rm Res}_{w_2^2\to 0} \M_5{}_{|z \to \bar R_2}  \bigg ] 
     \nnl & 
   e^{i b_2 k + i z (\hat u_2 k) b} \bigg [  
     {\rm Res}_{w_1^2\to 0 } \M_5{}_{|z \to R_1}  
     +  {\rm Res}_{w_1^2\to 0} \M_5{}_{|z \to \bar R_1}  \bigg ] 
     \bigg \}
. \end{align}
``
Here, the notation $|_{z \to R_2}$ ($|_{z \to R_1}$) amounts to setting the values of $z_1, z_2, z_v, z_b$ to  
$z_1 = - \frac{(\hat{u}_1 k)}{\sqrt{\gamma^2-1}}$, $z_2 = \frac{\gamma (\hat{u}_1 k)}{\sqrt{\gamma^2-1}}$ and $z_v = i s_\omega (\hat u_1 k) \sqrt{z^2+1}$, $z_b = (\hat u_1 k ) z$ ($z_v =  i s_\omega (\hat u_2 k) \sqrt{z^2+1} - \tilde v k$, $z_b=(\hat u_2 k)z - \tilde b k$),  and $s_\omega$ stand for the sign of $\omega$. Similarly, $|_{z \to \bar R_2}$ ($|_{z \to \bar R_1}$) is analogous, 
but with 
$z_v = - i s_\omega (\hat u_1 k) \sqrt{z^2+1}$,  ($z_v =  -i s_\omega (\hat u_2 k) \sqrt{z^2+1} - \tilde v k$). 
In this case,  $s_\omega$ can be effectively ignored because of the summation in the square bracket. 
We abbreviate the above formula as
\begin{align}
\label{eq:Int_zbRep} 
  I &=  
  - \frac{\pi}{\sqrt{\gamma^2-1}} \int_{-\infty}^\infty dz \frac{ \ e^{i b_1 k + i z (\hat{u_1 k) b}}}{\sqrt{z^2+1}}  \mathcal{R} \Big [R \Big ] + (1 \leftrightarrow 2), 
\end{align}
where $R \equiv  {\rm Res}_{w_2^2\to 0 } \M_5{}_{|z \to R_2}$. 
Since the residue can be expressed as a function of $w_2$ (after eliminating $w_1 \to k - w_2$), one can also equivalently write 
\begin{align} 
 R =  {\rm Res}_{w_2^2\to 0 } \M_5|_{w_2^\mu \to (\hat u_1 k) [\gamma \hat u_2^\mu - \hat u_1^\mu + z \tilde b^\mu + i \sqrt{z^2+1} \tilde v^\mu]  }
. \end{align} 

One could also choose to first integrate over $z_b$ in~\cref{Int_zvzb},  which leads to the $z_v$ representation. 
In this representation we are required to close the contour in the upper half-plane and the exponential ensures vanishing of the contribution of the big circle. 
Picking up the poles at 
$z_b = i \sqrt{z_v^2 + (\hat u_1 k)^2}$ and 
$z_b = i \sqrt{(z_v+ v k )^2 + (\hat u_2 k)^2}$ and putting the result in a form where the 
$1 \leftrightarrow 2$ symmetry is evident, we have
\begin{align}
\label{Int_zvRep}
I = - \frac{\pi}{\sqrt{\gamma^2-1}} \int_{-\infty}^\infty 
    \frac{dz}{\sqrt{z^2+1}}  e^{ib_1 k - s_\omega b (\hat u_1 k) \sqrt{z^2+1}} R
    + (1 \leftrightarrow 2)  
, \end{align}
where in this context
\begin{align} 
 R =  {\rm Res}_{w_2^2\to 0 } \M_5|_{w_2^\mu \to 
 (\hat u_1 k) [ \gamma \hat u_2^\mu - \hat u_1^\mu + i s_\omega \sqrt{z^2+1} \tilde b^\mu + z \tilde v^\mu]  }
. \end{align} 

The advantage of the $z_v$ representation is that the integral over $z$ in~\cref{Int_zvRep} is always convergent, for any sign of $\omega$,  thanks to the exponential suppression factor at large $z$. 
Thus, spectral waveforms can be evaluated numerically using this representation when an analytic computation is not possible. 
Conversely, the integral in~\cref{eq:Int_zbRep} may not always be convergent, and should then be interpreted in the distribution sense. 
The divergences correspond to $\delta(b)$ and its derivatives, which do not affect classical phenomena. 
They are dripped after inverse Fourier transformation of the spectral waveform to the time domain. 
On the other hand, the advantage of the $z_b$ representation is that the $d \omega$ integral for the transformation to the time domain is straightforward.  
In any case, both representations leads to exactly the same waveforms in the time domain. 

\section{More waveforms}
\label{app:MWF}

In this appendix we collect the spectral waveforms, from which the time domain waveforms displayed in the main text are calculated. 

We start with the gravitational waveforms in GR. 
At the zeroth order in spin, 
 \begin{align}  \hspace{-2 cm}
f_h^{{\rm GR}(0)}   =  &    
   -  { m_1 m_2 
     \over 512 \pi^2   \mpl^3   (\hat u_1 n)  \sqrt{\gamma^2 -1}   } 
      \int  {d z        e^{i \omega [ b_1 n +  z (\hat u_1 n) b ] } \over \sqrt {z^2+1} }  
     {\cal R} \bigg \{   { \big (\mathcal{F}_1^-(z) \big )^2 + \big (\mathcal{F}_2^-(z) \big )^2    
                \over  
       \gamma    (\hat u_2 n) -  (\hat u_1 n) + z (\tilde b n) + i \sqrt {z^2+1} (\tilde v n)       } \bigg \} 
       \nnl & 
    +  (1 \leftrightarrow 2) 
.   \end{align}   
At the first order in spin 
  \begin{align} \hspace{-2 cm}  
f_h^{{\rm GR}(1)}    =  &   
   { m_1 m_2 \omega  
     \over 512 \pi^2   \mpl^3   \sqrt{\gamma^2 -1}   } 
      \int  {d z   e^{i \omega [ b_1 n +  z (\hat u_1 n) b ] } \over \sqrt {z^2+1} }  
        {\cal R} \bigg \{   { (a_1 n) \over  (\hat u_1 n) } 
                  { \big (\mathcal{F}_1^-(z) \big )^2 + \big (\mathcal{F}_2^-(z) \big )^2
                \over  
       \gamma    (\hat u_2 n) -  (\hat u_1 n) + z (\tilde b n) + i \sqrt {z^2+1} (\tilde v n)      }
\nnl \hspace{-2 cm}   & 
   + (a_1^\mu + a_2^\mu)   
         \big ( \gamma \hat u_2^\mu - \hat u_1^\mu  +  z \tilde b^\mu + i \sqrt{z^2+1} \tilde v^\mu   \big ) 
{ \big (\mathcal{F}_1^-(z) \big )^2 - \big (\mathcal{F}_2^-(z) \big )^2 
         \over  
       \gamma    (\hat u_2 n) -  (\hat u_1 n) + z (\tilde b n) + i \sqrt {z^2+1} (\tilde v n)       }            
\nnl  \hspace{-2 cm}  &
+ {2  \mathcal{F}_1^-(z) \mathcal{F}_3^-(z)    \over    \gamma    (\hat u_2 n) -  (\hat u_1 n) + z (\tilde b n) + i \sqrt {z^2+1} (\tilde v n)    }  \bigg \}   +  (1 \leftrightarrow 2)  
 ,  \end{align}   
where we used the notation defined in \cref{F_functions_definition}

Moving to the scalar waveforms with matter conformally coupled to the scalar. 
At the zeroth order in spin, 
\begin{align}  
 f_\phi^{(0)}  =   & 
  - {  m_1 m_2  \over 64 \pi^2  \mpl^3    (\hat u_1 n)   \sqrt{\gamma^2- 1}  }  
    \int  { d z    e^{i \omega[ b_1 n +   (\hat u_1 n)  b z] }     \over \sqrt{ z^2 + 1} }  
\re    \bigg \{ 
 {  2 (\gamma^2-1)  \big [  c_2 (\hat u_1 n)^2  +  c_1 (\hat u_2 n)^2    \big ]     
 \over
 \gamma  (\hat u_2 n) - (\hat u_1 n)  +  (\tilde b n)  z + i (\tilde v n) \sqrt{ z^2 + 1}    } 
         \nnl  &  
 +   c_1 {  ( 2  \gamma^2   - 1 ) z (\tilde b  n)  
-  (\hat u_1 n) 
- \gamma(2 \gamma^2 - 3 )  (\hat u_2 n)  
   \over   \gamma^2-1  } 
         + C_1^{(0)} c_2    (\hat u_1 n)   
\bigg \}    + (1 \leftrightarrow 2 )
,      \end{align}   

At the first order in spin, 
   \begin{align} 
   \hspace{-2.6cm} 
  f_{\phi}^{(1)}  = & 
  - {i m_1 m_2  \omega  \over 32  \pi^2 \mpl^3  }
\int  {d z e^{i \omega [ b_1 n + z (\hat u_1 n) b ]  }  \over \sqrt {z^2+1} }     \re  \bigg \{
 c_1  \big [ z (\tilde v n)   - i \sqrt{z^2+1}  (\tilde b n)    \big ]  
  \big [  - (\hat u_1 a_2)  +  z (\tilde b a_2) + i \sqrt{z^2+1} (\tilde v a_2)  \big ]   
 \nnl   \hspace{-2.6cm}  \times & 
   \bigg [  { \gamma  \over \gamma^2-1 }  
-     {    (\hat u_2 n)   \over 
 \gamma    (\hat u_2 n) -  (\hat u_1 n) + z (\tilde b n) + i\sqrt {z^2+1} (\tilde v n)   }  \bigg ] 
+       {  c_2   (\hat  u_1 n)    \over  
 \gamma    (\hat u_2 n) -  (\hat u_1 n) + z (\tilde b n) + i \sqrt {z^2+1} (\tilde v n)   } 
 \nnl   \hspace{-2.6cm}  \times &  
         \bigg [
 \big [    z   (\tilde v n) - i \sqrt{z^2+1}  (\tilde b n)   \big ] (\hat u_2 a_1) 
- \big [  \gamma  (\tilde v n)  +  i  \sqrt{z^2+1}  \big ( \gamma (\hat u_1 n)  - \hat u_2 n \big )     \big ]  (\tilde b a_1)  
-   \big [   z    \big ( \hat u_2  n  - \gamma (\hat u_1 n) \big )   -  \gamma  (\tilde b n)    \big ] (\tilde v a_1)   
\bigg ]   
\nnl  \hspace{-2.6cm}  &  
+ {C_1^{(1)}  c_2  \over 2 \sqrt{\gamma^2- 1}}
\bigg [   \gamma (\hat u_1 n)  (\hat u_2 a_1)  +  z(\hat u_1 n)  (\tilde b a_1) - (a_1 n) \bigg ]
\bigg \} 
 + (1 \leftrightarrow 2) 
  .  \end{align}

\section{More amplitudes} 
\label{app:higheramps}

In this appendix we give the full expressions for the quantum four-point amplitudes whose classical limit we presented in~\cref{sec:Four-point_amplitudes}. We will also give the expression for a contribution to the four-point amplitudes and the five-point residues proportional to $\mathcal{O}(c_n \hat{\alpha})$, which can appear in SGB/DCS theories and give an argument on why this contribution can be neglected in our computations.

\subsection{Full four-point amplitudes}
\label{app:fourpoint}

We start with the 2 massive - 2 scalars four-point amplitude in~\cref{Amp_Phi-bar{Phi}-phi-phi}. The full quantum amplitude reads:

\begin{align}
    \M_{U}\big [1_{\Phi}, 2_{\bar{\Phi}}, 3_\phi, 4_\phi\big ] = - \frac{1}{2 \mpl^2 m^{2S} s} \bigg \{ &(t-m^2)^2 (\langle \bm 2 \bm 1\rangle^{2S} + [\bm 2\bm 1]^{2S} )\nnl
	& + (t- m^2) [ \langle \bm2|p_3 q|\bm 1\rangle +[ \bm 2|p_3 q|\bm 1]  ] \sum_{k=0}^{2S-1} \langle \bm 2 \bm 1\rangle^{k} [\bm 2\bm 1]^{2S-1-k} \bigg \} + \mathcal{O}(c_n^2)
. \end{align}

We now move to the amplitude with 1 graviton and 1 scalar. Here there is a contribution from the conformal coupling, see \cref{Amp_Phi-bar{Phi}-phi-h}, which away from the classical limit reads
\begin{align}
    \M_U \big [1_{\Phi}, 2_{\bar{\Phi}}, 3_\phi, 4_h^- \big ] =  \frac{c_n m^2 }{\mpl^2} \frac{[\bm 2\bm 1]^{S}}{m^{2S}}  \bigg \{ &(-1)^S \frac{\langle 4|p_3p_1|4 \rangle^2 \langle \bm 2 \bm 1\rangle^S }{s (t-m^2) (u-m^2) } \nnl
	&-  (-1)^{S-1} \frac{\langle 4|p_3p_1|4 \rangle^2 \langle 4 \bm1 \rangle \langle 4 \bm2 \rangle  \langle \bm 2 \bm 1\rangle^{S-1} }{(t-m^2)(u-m^2)} \nnl
	& - \frac{\langle 4 \bm1 \rangle^2 \langle 4 \bm2 \rangle^2}{t-m^2} \sum_{k=2}^S \binom{S}{k} (-1)^{S-k} \frac{\langle \bm 2 \bm 1\rangle^{S-k}  \langle \bm 1|p_4 |\bm 2]^{k-2}}{m^{k-2}} \nnl
	&- \frac{\langle 4 \bm1 \rangle^2 \langle 4 \bm2 \rangle^2}{u-m^2} \sum_{k=2}^S \binom{S}{k} (-1)^{S-2k} \frac{\langle \bm 2 \bm 1\rangle^{S-k} \langle \bm 2|p_4 |\bm 1]^{k-2}}{m^{k-2}} \bigg \},
 \end{align}
and $ \M_U\big [1_{\Phi}, 2_{\bar{\Phi}}, 3_\phi, 4_h^+ \big ]$ can be easily derived using crossing symmetry. 

\label{app:Full_Four-points}

\subsection{Subleading amplitudes}
\label{app:Subleading_Amp_Res_SGB/DCS}

If we only allow for the scalar conformal coupling, as well as its minimal coupling to gravity,  the amplitudes in~\cref{app:fourpoint} and associated contact terms are the only beyond-GR effects relevant for calculating leading PM radiation.
However, in the SGB/DCS framework, the scalar $\phi$ also couples to GB/CS invariants with the coupling denoted as $\hat \alpha$. 
This leads to additional $\cO(c_n \hat \alpha)$ and $\cO(\hat \alpha^2)$ corrections to the amplitudes.  
Since in the SGB framework $c_n \sim \hat \alpha$, these are of the same order in $\hat \alpha$ as the effects of the monopole scalar charge calculated in the main text. 
However, it turns out that they are subleading in the classical limit.  
It was already shown in Ref.~\cite{Falkowski:2024bgb} that the $\cO(\hat \alpha)$ contributions to 
$\M_{U}\big [1_{\Phi}, 2_{\bar{\Phi}}, 3_\phi, 4_h \big ]$ are 
$\cO(\hbar^2)$ in the classical scaling. 
Below we demonstrate the same holds for the $\cO(c_n \hat \alpha)$ contributions to the Compton amplitude 
$\M_{U}\big [1_{\Phi}, 2_{\bar{\Phi}}, 3_h, 4_h \big ]$. 
Indeed, the graviton cut contribution to that amplitude leads to  
\begin{align}
\Delta \M_{U}\big [1_{\Phi}, 2_{\bar{\Phi}}, 3_h^-, 4_h^- \big ] = \frac{2c_n \hat{\alpha} m^2}{\Lambda^2 \mpl^2} \frac{\langle 3 4 \rangle^4}{s} \frac{\langle \bm 2 \bm 1\rangle^S [ \bm 2 \bm1]^S}{m^{2S}}
. \end{align}
The other helicity configuration is obtained by crossing symmetry.
In the classical limit all the spin dependence becomes trivial:
\begin{align}
    \M^{\rm cl}_{U}\big [1_{\Phi}, 2_{\bar{\Phi}}, 3_h^-, 4_h^- \big ] &= \frac{2c_n \hat{\alpha} m^2}{\Lambda^2 \mpl^2} \frac{\langle 3 4 \rangle^4}{s} , \nnl
    \M^{\rm cl}_{U}\big [1_{\Phi}, 2_{\bar{\Phi}}, 3_h^+, 4_h^+ \big ] &= \frac{2c_n \hat{\alpha}^* m^2}{\Lambda^2 \mpl^2} \frac{[3 4 ]^4}{s}  
    \label{Amps_mod_grav_Compton}
. \end{align}
 
This results is $\cO(\hbar^2)$ in the classical scaling, as advertized, 
compared to $\cO(\hbar^0)$ for the $\cO(c_n^2)$ contributions found in \cref{sec:Four-point_amplitudes}. 
Therefore, the corrections due to \cref{Amps_mod_grav_Compton} are quantum in the popular parlance, are non-resolvable in the sense of Ref.~\cite{Bellazzini:2022wzv}.
Physically, this means the associated contributions to the waveforms are suppressed by  two powers of the small dimensionless parameter ${1 \over b m_n}$ compared to those found in~\cref{sec:waveforms}.  
By the same argument, the effects calculated in Ref.~\cite{Falkowski:2024bgb} are non-resolvable whenever the monopole scalar charge $\cO(c_n^2)$ contributions are present. 
They can be phenomenologically relevant only in scalar-tensor theories which do not admit hairy solutions with $c_n \sim \hat \alpha$.

Finally we comment that are no contact terms in the Compton amplitude up to linear order in spin,  as it is shown in~\cref{app:CONTACT}. The first contact term arrives at quadratic order and can be written as:
\begin{align}
     \M^{\text{ cl}, (2)}_{C} \big [1_{\Phi}, 2_{\bar{\Phi}}, 3_h^-, 4_h^-\big ] &=  \frac{\hat{F}^{(2)}_1 m^2}{\Lambda^2 \mpl^2} \langle 3 4 \rangle^4 a_1^2 \nnl
     \M^{\text{ cl}, (2)}_{C} \big [1_{\Phi}, 2_{\bar{\Phi}}, 3_h^+, 4_h^+\big ] &=  \frac{\hat{F}^{(2)*}_1 m^2}{\Lambda^2 \mpl^2} [34]^4 a_1^2
. \end{align}

\section{Classical contact terms} 
\label{app:CONTACT}

In this appendix we will discuss how to construct the general contact terms for all four-point amplitudes we considered. This approach follows similar steps as in appendix B of~\cite{Falkowski:2024bgb} and we refer the reader to that reference for a more detailed explanation of the idea of this procedure. Nevertheless, we extend the discussion in that reference to account for the new amplitudes we need to consider in this case, which exhibit Bose symmetry.

We start with some definitions, which will be relevant for the next sections. Namely, we assume an amplitude containing two massive ($1$ and $2$) with mass $m$ and two massless particles ($3$ and $4$). We denote their momenta $p_1, p_2, p_3, p_4$ and the mandelstam invariants $s = 2 p_3 p_4, \ t-m^2 = 2 p_1 p_3 , \ u-m^2= 2 p_1 p_4$. Unless both massless particles are scalars, we can define a family of polarization vectors as:

\begin{align}
    \epsilon^\mu_{i,-} = \frac{\langle i | \sigma^\mu | \tilde{\zeta}]}{\sqrt{2}[i \tilde{\zeta}]} \qquad \qquad , \qquad \qquad \epsilon^{\mu}_{i,+} = \frac{\langle \zeta | \sigma^\mu | i]}{\sqrt{2}\langle i \zeta \rangle } ,
\end{align}
where $i=3,4$ and for the construction we always choose $| \zeta \rangle = |4 \rangle$ and $| \tilde{\zeta} ] = | 4 ]$ if $i=3$ and vice versa. We introduce the normalized spin vector as $a_1^\mu = \frac{s_1^\mu}{m}$, where $s_1^\mu$ is the classical expectation value of the Pauli-Lubanski vector for particle $1$. We also define the optical parameter $\xi$ as:

\begin{align}
\xi^{-1} \equiv   \frac{m^2 s}{(t-m^2)(u-m^2)} ,
\end{align}
and the $w^\mu$, $\tilde{w}^\mu$ vectors\footnote{note that we can express these vectors with respect to either the polarization vectors of particle $3$ or $4$, if both are non-scalar particles, here we just make a choice}:

\begin{align}
    w^\mu= \frac{(t-m^2)}{2(p_1 \epsilon_{3,-})} \epsilon_{3,-}^\mu \qquad \qquad , \qquad \qquad \tilde{w}^\mu = \frac{(t-m^2)}{2(p_1 \epsilon_{3,+})} \epsilon_{3,+}^\mu ,
\end{align}

Lastly, to capture non-conservative effects we include in our analysis:

\begin{align}
    |a_1|'=\frac{(t-m^2)}{2m} |a_1| .
\end{align}

The properties and $\hbar$ scaling of these quantities are discussed in detail in Ref.~\cite{Falkowski:2024bgb}. In that context, assuming that the amplitude does not need to satisfy additional properties, a basis for the construction of contact terms can be the same as in that reference: $\{ p_3 \cdot a_1, p_4 \cdot a_1, w \cdot a_1, |a_1|' \}$\footnote{Note that all the operators in this basis have to be dimensionless and scale as $\sim \hbar^0$}. 

However, we will later see that in this paper we will also be interested in building contact terms for the amplitudes describing scattering of scalars, $ \M^{\rm cl} \big [1_{\Phi}, 2_{\bar{\Phi}}, 3_\phi, 4_\phi\big ]$, and gravitons for the helicity flipping case, $ \M^{\rm cl} \big [1_{\Phi}, 2_{\bar{\Phi}}, 3_h^s, 4_h^s \big ]$. For these amplitudes we should also impose Bose symmetry, i.e. the amplitude should not change under $3 \leftrightarrow 4$ exchange. We can quickly see that modifying the basis mentioned before in such a way such that it respects this symmetry is not straightforward. We can check that it is in fact impossible modify $w^\mu$ such that it remains a vector related to the polarization vectors, scales as $\hbar$ and also respects the Bose symmetry\footnote{For the case of the scattering of 2 scalars there is in fact no polarization vectors at all, but we try to keep the discussion as general as possible.}. The same is true for the $|a_1|'$ term we considered. The reason is quite simple: In order to impose the Bose symmetry one should make the kinematic part of each structure $3 \leftrightarrow 4$ symmetric, which translates to $(t-m^2) \to (t-m^2 + u-m^2) \sim \mathcal{O}(\hbar^2)$, since $t-m^2 \approx - (u-m^2)$ at leading order in the classical limit \footnote{For the $w^\mu$ vector we should in principle also make the change $\frac{\epsilon_{3,-}^\mu}{2(p_1 \epsilon_{3,-})}  \to \Big ( \frac{\epsilon_{3,-}^\mu}{2(p_1 \epsilon_{3,-})} + \frac{\epsilon_{4,-}^\mu}{2(p_1 \epsilon_{4,-})} \Big )$ in order to respect the symmetry, but this doesn't affect our conclusion.}. Therefore, these structures will only contribute to subleading $\hbar$ orders. Notice that this means that \emph{imposing Bose symmetry doesn't allow for dissipative effects at leading order}.

Still we have to be careful as we cannot just completely neglect the $w^\mu$ vector that simply. The reason is that, in the classical limit, one can construct a polarization vector-independent quantity through this vector~\cite{Bautista:2022wjf,Falkowski:2024bgb}:

\begin{align}
 \frac{(t-m^2)(u-m^2)}{4 m^2} a_1^2 \approx \xi \big ( w\cdot  a_1 - p_3 \cdot a_1 \big ) \big ( w \cdot a_1 + p_4 \cdot a_1 \big ) + \big (w \cdot a_1)^2 . 
\end{align}

Thus, we include the $a_1^2$ term, dressed with an overall $s$ in front to fix dimensions and $\hbar$ scaling. In total, in these cases our basis will read:

\begin{align}
  \bigg \{ \quad    p_3 \cdot a_1 + p_4 \cdot a_1 \quad , \quad   s a_1^2   \quad  \bigg \} .\label{basis_for_cont}
\end{align} 

\subsection{2 massive - 2 scalars}

We first discuss the 2 massive - 2 scalars amplitudes. We use the basis in~\cref{basis_for_cont} and present the results for the contact terms' construction starting with the amplitude describing the scattering of scalars: 
\begin{align}
    \M^{\rm cl}_{}\big [1_{\Phi}, 2_{\bar{\Phi}}, 3_\phi, 4_\phi\big ]  &=  \frac{1}{\mpl^2 }  \frac{(t-m^2) (u-m^2)}{s} \bigg \{  \cosh (q \cdot a_1) + 2i  \frac{\varepsilon_{\mu \nu \rho \sigma} p_1^\mu p_3^\nu p_4^\rho a_1^\sigma}{u-m^2} \frac{\sinh (q \cdot a_1)}{q \cdot a_1} + P_{1,\xi} \bigg \} \nnl
    &=  \frac{m^2}{\mpl^2 } \xi  \bigg \{  \cosh (q \cdot a_1) + 2i  \frac{\varepsilon_{\mu \nu \rho \sigma} p_1^\mu p_3^\nu p_4^\rho a_1^\sigma}{u-m^2} \frac{\sinh (q \cdot a_1)}{q \cdot a_1} + P_{1,\xi} \bigg \} ,
\end{align}
where $P_{1,\xi}$ denotes a polynomial in our basis \cref{basis_for_cont}, which is simultaneously a Laurent expansion in $\xi$. This polynomial is constructed as such to include all classical contact terms' deformations. We can immediately notice a feature not encountered before: At $\xi^{-1}$ order in the Laurent expansion, we can actually cancel fully the pole dependence without altering the $\hbar$ behavior of the amplitude. This indicates the existence of a \emph{classical contact term at zeroth order in the spin}. The polynomial expanded around $\xi^0$ takes the form:

\begin{align}
	P_{1, \xi} =... &+ \xi^{-1} \sum_{ij} A^{-1}_{ij}  	\Big [ p_3 \cdot a_1 + p_4 \cdot a_1 \Big ]^i \big [s a_1^2 \big ]^j  \nnl
	&+ \xi^0 (s a_1^2) \sum_{ij} A^{0}_{ij}  	\Big [ p_3 \cdot a_1 + p_4 \cdot a_1 \Big ]^i \big [s a_1^2 \big ]^j \nnl
	& + \xi^{1} (s a_1^2)^2 \sum_{ij} A^{1}_{ij}  	\Big [ p_3 \cdot a_1 + p_4 \cdot a_1 \Big ]^i \big [s a_1^2 \big ]^j +... , 
\end{align}
where the sums runs from $0$ and are truncated at an integer such that there are no more than $2S$ insertions of the spin vector in each term. The zeroth order in the spin term corresponds to the $i=j=0$ term multiplying the $\xi^{-1}$ term and leads to the contact term in~\cref{Amp_Phi-bar{Phi}-phi-phi_Contact0}. At linear order in the spin, we can again check that the only contributions come from the $\xi^{-1}$ part of the expansion, namely the $i=1, \ j=0$  part, which we wrote down in~\cref{Amp_Phi-bar{Phi}-phi-phi_Contact1}.
\label{app:2mass2scal}

\subsection{2 massive - 1 scalar - 1 graviton}

\label{app:2mass1scal1grav}

We now move on to the case of the amplitude containing 1 scalar and 1 graviton. The treatment is essentially the same as in Ref.~\cite{Falkowski:2024bgb}. The full amplitude, including contact terms, can be read as:

\begin{align}
    \M^{\rm cl} \big [1_{\Phi}, 2_{\bar{\Phi}}, 3_\phi, 4_h^-\big ] &=  \frac{2 c m^2}{ \mpl^2} \frac{s (p_1 \epsilon_-)^2 }{(t-m^2)(u-m^2) } \bigg \{ 1 + P_{3, \xi} \bigg \} \nnl 
	&=  \frac{2 c}{ \mpl^2} \xi (p_1 \epsilon_-)^2 \bigg \{ 1 + P_{3, \xi} \bigg \} ,
\end{align}
and around $\xi^0$ $P_{3,\xi}$ takes the form:

\begin{align}
	P_{3,\xi} = ... &+ \xi^{-1} (wa_1 + p_4 a_1)^2 \bigg \{ \sum_{ijk} F^{-1}_{ijk} (p_3 a_1 )^i (p_4 a_1)^j (wa_1)^k \bigg \} \Big (1 + F^{-1} |a_1|' \Big ) \nnl
	&+ \xi^0 (wa_1)^2 (wa_1+p_4 a_1) \bigg \{  \sum_{ijk} F^{0}_{ijk} (p_3 a_1 )^i (p_4 a_1)^j (wa_1)^k \bigg \} \Big (1 + F^{0} |a_1|' \Big ) \nnl
	&+ \xi^0 (wa_1) |a_1|' (wa_1+p_4 a_1)  \bigg \{  \sum_{ijk} G^{0}_{ijk} (p_3 a_1 )^i (p_4 a_1)^j (wa_1)^k \bigg \} \Big (1 + G^{0} |a_1|' \Big ) \nnl
	&+ \xi^1 (wa_1)^4 \bigg \{  \sum_{ijk} F^{1}_{ijk} (p_3 a_1 )^i (p_4 a_1)^j (wa_1)^k \bigg \} \Big (1 + F^{1} |a_1|' \Big ) \nnl
	&+ \xi^1 (wa_1)^3 |a_1|' \bigg \{  \sum_{ijk} G^{1}_{ijk} (p_3 a_1 )^i (p_4 a_1)^j (wa_1)^k \bigg \} \Big (1 + G^{1} |a_1|' \Big ).
\end{align}

Observe that the first contact term corrections come at quadratic order in the spin from the $\xi^{-1}$ term. Thus, we correctly didn't include any contact terms' deformations up to linear order in the spin in our analysis. The polynomial for a positive helicty graviton can be found using crossing symmetry.

\subsection{2 massive - 2 gravitons}

The last case to consider is the correction to the helicity flipping gravitational Compton amplitude which we examined in~\cref{app:Subleading_Amp_Res_SGB/DCS}:

\begin{align}
	  \M^{\rm cl} \big [1_{\Phi}, 2_{\bar{\Phi}}, 3_h^-, 4_h^-\big ] &= \frac{2c \hat{\alpha} m^2}{\Lambda^2 \mpl^2} \frac{\langle 3 4 \rangle^4}{s} \bigg \{ 1 + P_{4, \xi} \bigg \} \nnl
	& = \frac{2c \hat{\alpha} m^2}{\Lambda^2 \mpl^2} \frac{\langle 3 4 \rangle^3}{ [ 43] } \bigg \{ 1 + P_{4, \xi} \bigg \} , \label{Amps_mod_grav_Compton_Contact}
\end{align}
where $P_{4,\xi} $ around  $\xi^0$ is given by:

\begin{align}		
	P_{4, \xi} =& ...+ \xi^0 s a_1^2 \sum_{ij}  \Theta^{0}_{ij} (p_3 a_1 + p_4 a_1)^i (sa^2)^j \nnl
		& + \xi^{1} (s a_1^2)^2 \sum_{ijk}  \Theta^{1}_{ijk} (wa_1)^i (p_3 a_1)^j (p_4 a_1)^k (1 + \Theta^{1} |a_1|') +... \ ,
\end{align}
again using the basis in~\cref{basis_for_cont} and it is implicit that all terms in this sum are built such that they obey the symmetry under $3 \leftrightarrow 4$ exchange. There are in fact no contact terms at $\xi^{-1}$ orders or lower, because it is not possible to cancel poles in the $t$ and $u$ channels. The first contact terms' corrections come at quadratic spin order

Attention has to be drawn that \emph{this analysis is not the same from the one for the gravitational Compton amplitude coming from minimally coupled gravity}. The reason is that, when considering modifications to gravity (or any other theory for that matter) one has to take into account the 1) pole structure and 2) the overall $\hbar$ counting, when constructing classical contact terms. It's evident that both are different from the helicity-flipping Compton amplitudes in minimally coupled gravity. 

Note that the difference in the $\hbar$ scaling between this amplitude and the amplitudes including higher order effects in $\hbar$ has to do with extra derivatives acting on the massless fields as is the case for example in Ref.~\cite{Falkowski:2024bgb}. This translates to the extra $\hbar$ factors in front of the amplitudes, which could come for example in the form of the mandelstam invariants. These can be canceled when including contact terms, leading to potentially a set of contact terms at lower orders in the spin than initially expected. 

\label{app:Contact_2mass_2grav}


\bibliographystyle{JHEP}
\bibliography{refs} 

\end{document}